\documentclass[twocolumn]{article}

\usepackage{scrextend}
\usepackage{lineno,hyperref}
\usepackage{threeparttable}
\usepackage{subcaption}
\usepackage{graphicx}
\usepackage{adjustbox}
\usepackage{tablefootnote}
\usepackage{amsmath}
\usepackage[affil-it]{authblk}
\usepackage{xcolor}

\begin{document}

\title{\vspace{-2.0cm}\textbf{GENETIC EVOLUTION OF A MULTI-GENERATIONAL POPULATION IN THE CONTEXT OF INTERSTELLAR SPACE TRAVELS} 
\\ \vspace{0.3cm} \textbf{\Large Part I: Genetic evolution under the neutral selection hypothesis}}
   
\author{Fr\'ed\'eric Marin\textsuperscript{1}, Camille Beluffi\textsuperscript{2} \& Fr\'ed\'eric Fischer\textsuperscript{3}\\
{\small 1 Universit\'e de Strasbourg, CNRS, Observatoire astronomique de Strasbourg, UMR 7550, F-67000 Strasbourg, France\\
2 CASC4DE, Le Lodge, 20, Avenue du Neuhof, 67100 Strasbourg, France\\
3 Institut de physiologie et de chimie biologique, Laboratoire Dynamique \& Plasticit\'e des Synth\'etases, UMR 7156, F-67000 Strasbourg, France}}
              
\date{Dated: \today}

\twocolumn[
  \begin{@twocolumnfalse}
    \maketitle
    \vspace{-1\baselineskip}
    \begin{abstract}
    We updated the agent based Monte Carlo code HERITAGE that simulates human evolution within restrictive 
    environments such as interstellar, sub-light speed spacecraft in order to include the effects of population
    genetics. We incorporated a simplified  -- yet representative -- model of the whole human genome with 46 
    chromosomes (23 pairs), containing 2110 building blocks that simulate genetic elements (loci). Each individual
    is endowed with his/her own diploid genome. Each locus can take 10 different allelic (mutated) forms that can be 
    investigated. To mimic gamete production (sperm and eggs) in human individuals, we simulate the meiosis process 
    including crossing-over and unilateral conversions of chromosomal sequences. Mutation of the genetic information
    from cosmic ray bombardments is also included. In this first paper of a series of two, we use the neutral
    hypothesis: mutations (genetic changes) have only neutral phenotypic effects (physical manifestations), implying 
    no natural selection on variations. We will relax this assumption in the second paper. Under such hypothesis, we
    demonstrate how the genetic patrimony of multi-generational crews can be affected by genetic drift and mutations. 
    It appears that centuries-long deep space travels have small but unavoidable effects on the genetic
    composition/diversity of the traveling populations that herald substantial genetic differentiation on longer 
    time-scales if the annual equivalent dose of cosmic ray radiation is similar to the Earth radioactivity background 
    at sea level. For larger doses, genomes in the final populations can deviate more strongly with significant genetic 
    differentiation that arises within centuries. We tested whether the crew reaches the Hardy-Weinberg equilibrium
    that stipulates that the frequency of alleles (for non-sexual chromosomes) should be stable over long periods. 
    We demonstrate that the Hardy-Weinberg equilibrium is reached for starting crews larger than 100 people, confirming
    our previous results, while noticing that larger departing crews (500 people) show more stable equilibriums over time.  
    \end{abstract}    
    
    \vspace{3\baselineskip}
  \end{@twocolumnfalse}
]

\section{Introduction}
\label{Introduction}

Why go explore exoplanets? One of the fundamental purposes of space exploration is the search for life other than the one we know 
on Earth. This question has obsessed humanity for thousands of years and is even found in ancient philosophical writings (Thales, 
Anaximander, Bruno, Kant ...). The discovery of life on an extraterrestrial world within our Solar System, on an exoplanet or on
an exomoon would, of course, be of prime interest. This would allow us to answer the fundamental question as to whether we (humans, 
and all other life forms found on Earth) are the only living beings in the Universe. In addition, such a discovery could also help 
better understand abiogenesis, that is the ``natural generation'' of life from non-living matter. Indeed, terrestrial life is based 
on a (bio)chemistry that involves only a few atoms: carbon, hydrogen, nitrogen, oxygen, phosphorus and sulfur (usually known through 
the mnemonic acronym \textit{CHNOPS}). All those atoms are used by cells to construct simple to extremely complex high molecular 
mass molecules. Other trace elements (iron, zinc, potassium, sodium, boron, etc.) are also crucial to all living forms, although 
they are not integral part of macromolecules, but rather essential to the functioning of these molecular machines. Organic matter 
(carbon-containing molecules) now appears to be widespread in the Universe, including complex kinds of polymers, especially within
meteorites \cite{McGeoch2015}, but the question of the origin of the first terrestrial complex molecular entities that underwent 
self-reproduction and used coordinated and complex chemical reaction networks (metabolism) to maintain their structural features remains
a mystery. Moreover, are other types of chemistries (other than carbon-based) possible? 

To answer these questions, satellites and rovers
have been sent all over the Solar System. The wealth of planets, moons, asteroids and comets now explored tells us that potential living 
pools that could be suitable to life as we know it (Europa, Titan, Mars, see \cite{Carr1998}, \cite{Iess2012}, and \cite{Orosei2018}, 
respectively) do exist. However, with the exception of Mars, none of the celestial bodies thus far explored within our Solar System seem 
to have kept for long enough periods of time conditions similar to those that presided on Earth when the first life forms are supposed to 
have emerged. Those facts make exoplanets and exomoons very interesting alternative candidates to study those questions. However, the 
tremendous distances between our planet and any exoplanet lead to missions taking centuries using non-fictional means of propulsion. This 
inevitably excludes deep-space exploration beyond the Solar System to first exploit resources for commercial or economic purposes, and rather
highlights that it would necessarily be for other, likely scientific, goals.

One of the immediate consequences of those distances is that crewed journeys cannot be achieved within the life expectancy of a human.
Discarding non-mature options (cryogenic technologies, suspended animation scenarios and genetic arks), the best choice might 
be to rely on giant self-contained generation ships that would travel through space while their population is active \cite{Hein2012}.
Such an undertaking requires choosing an initial crew in such a way 
that its overall genetic diversity be sufficient to sustain a long-term multi-generational voyage in an enclosed environment. Here,
genetic diversity refers to the amount of variations that are present on average within the population that would minimize inbreeding 
and consanguinity, under the enclosed and isolated conditions that the crew and all subsequent offspring will endure during the course
of space travel. Inbreeding and consanguinity have well-documented consequences on health \cite{Fareed2017} and fertility \cite{Charlesworth2009}. 
This directly impacts the population's genetic health, which constrains the choice of a minimal viable population (MVP) \cite{Jamieson2012,Frankham2013}. 
In addition, since migration back to or forward from Earth would be impossible given timescales and technological costs, the space-faring 
crew should be regarded as an ever after enclosed and henceforth isolated population, with no possible external genetic input. In this 
context, choosing an initial crew is complex because we need to introduce enough genetic variations in the spaceship’s population
to avoid the genetic diversity to decrease with time, to the extent that inbreeding and consanguinity eventually prevails.

In our agent based Monte Carlo code HERITAGE \cite{Marin2017,Marin2018,Marin2019,Marin2020}, because no genotype was formerly attributed 
to individuals, we took advantage of the precisely defined genealogy and kinship of individuals in the crew to evaluate the dynamics of 
consanguinity using the coefficients of inbreeding ($C_{\rm i}$) and of consanguinity ($F$) introduced by Wright \cite{Wright1922}. They 
take into account only relationships of father/mother couples to their common ancestor at the generations scale \cite{Wright1922}. 
The minimal number of initial crew members (male/female-balanced) that ensured $F$ to remain below a given security threshold expected not 
to be deleterious for the enclosed population was of $\sim$ 100 people to ensure a thousand year-long journey to Proxima Centauri~$b$ 
\cite{Marin2018}. As stated, $Ci$ and $F$ do not take into account genetic features of the initial crew that was presupposed to be genetically 
diverse (with low genetic similarities at the population level). However, because of a mechanism called
genetic drift, the genetic composition of a starting population has to be taken into account to more realistically evaluate the evolution of 
inbreeding over time in an enclosed environment. This is what Smith \cite{Smith2014} did when he applied population genetics probabilistic 
principles to determine a MVP of 14000 -- 44000 individuals to have a healthy group of settlers upon a 5 generation-long journey. To 
this end, he followed the principle of reduction in heterozygosity (ROH, that is a measure of genetic diversity) applied to one single genetic element. 
ROH has effects similar to consanguinity in terms of health and reproductive outcomes \cite{Ceballos2018} and accordingly affects the genetic health of a population \cite{Howard2017}. 

The discrepancies between Smith's and our results as to determine the MVP of the spaceship likely originates from the use of 
different methodologies that either integrate only kinship or simplified probabilistic population genetics principles that cannot
alone approximate the complexity of population genetics at the genome scale. We therefore reasoned that to understand, quantify,
evaluate and predict the complex genetic phenomena involved, we needed virtual populations in which individuals are endowed
with bona fide genetic features mimicking those found in human genomes to perform forward-in-time population genetics simulations.
Those are the goals of the present paper and its forthcoming accompanying publication (part~II).

\section{Adding genetics to HERITAGE}
\label{Improvements}

In order to include a representative model of the human genome and its evolution through multiple centuries of space travel -- 
that, in addition, follow the laws of heredity and genetics --, we gradually improved HERITAGE. In the following, we detail the 
many upgrades of HERITAGE that will allow us to build increasingly realistic initial populations to simulate genetic outcomes of 
space-faring demes\footnote{In biology, demes are considered small and randomly breeding groups of individuals that are collectively 
more likely to mate with one another than with any other individual that belongs to another deme \cite{Carson1987}.}. In 
Sec.~\ref{Improvements:chromosomes} we present the methodology to include chromosomes, loci and alleles in HERITAGE. In 
Sec.~\ref{Improvements:population_ini} we show how initial population genetics (the zeroth-generation) can be modeled. In 
Sec.~\ref{Improvements:meiosis} we describe the principles for transmitting the genetic traits from the parents to the offspring
through gamete production and meiosis. We use this improvement to show the impact of allelic gene crossing-over and conversion 
onto the population genomes before including the effects of mutations and cosmic ray radiation in Sec.~\ref{Improvements:mutations}
and Sec.~\ref{Improvements:cosmic}, respectively.

\subsection{Methodology}
\label{Improvements:chromosomes}

\subsubsection{The human genome: how to model it?}
\label{Improvements:chromosomes:Model}

As stated in the introduction, a genealogical approach to measure the degree of consanguinity of individuals is a good way to 
evaluate the genetic diversity and health at the population level. However, the degree of genetic relatedness -- that can lead to
consanguinity in the offspring over time in small populations -- strongly depends on the starting genetic composition of the initial 
population. It can be a continuum between high and low values, something that was not taken into account in our previous studies. In addition, 
the randomness of mating histories of and between individuals within genealogical lines can significantly modify the genetic 
composition (e.g., the frequency of alleles) of the overall descendant population throughout generations. This stochastically
changes the degree of genetic relatedness between individuals, a phenomenon that not only depends on these individual histories, but
also on the population's size at each step. In order to better simulate human populations during the course of the interstellar
journey by taking into account the genetic constitution of individuals and of the overall population, we needed to provide 
individuals with virtual genomes. 

\begin{figure*}[!t]
\centering
  \includegraphics[trim = 0mm 0mm 0mm 0mm, clip, width=\textwidth]{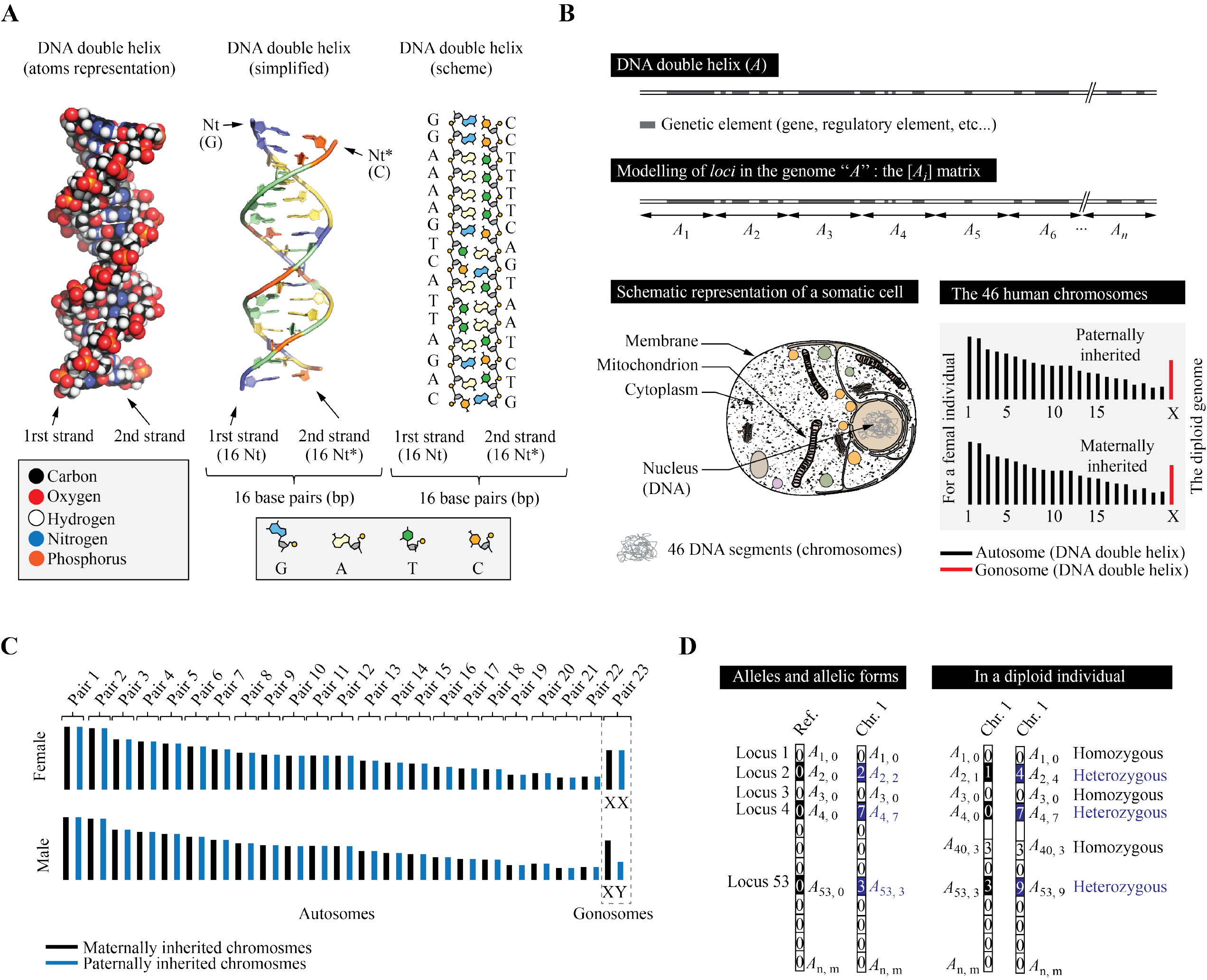}
  \caption{Model of the human genome such as implemented in HERITAGE. Description 
	   of the figure can be found in the text. Nt: (deoxyribo)nucleotide, 
	   Nt$^*$: complementary Nt. A somatic (non-sexual) cell is represented to 
	   emphasize the nuclear localization of chromosomes within cells 
	   (in the nucleus).}
  \label{Fig:Genome}
\end{figure*}

\begin{table*}[!t]
  \centering
  \begin{adjustbox}{width=\textwidth}
    \begin{tabular}{cccccc} 
      \textbf{Chromosome} & \textbf{Size (x10$^6$ base pairs)} & \textbf{Number of genes (N)} & \textbf{N/10} & \textbf{N/50} & \textbf{Simulated number of loci per chromosome}\\
      \hline
      1 & 248.96 & 5096 & 509.6 & 101.92 & 100 \\ 
      2 & 242.19 & 3867 & 386.7 & 77.34 & 70 \\ 
      3 & 198.3 & 2988 & 298.8 & 59.76 & 60 \\ 
      4 & 190.22 & 2438 & 243.8 & 48.76 & 50 \\ 
      5 & 181.54 & 2594 & 259.4 & 51.88 & 50 \\ 
      6 & 170.81 & 3014 & 301.4 & 60.28 & 60 \\ 
      7 & 159.35 & 2770 & 277 & 55.4 & 55 \\ 
      8 & 145.14 & 2170 & 217 & 43.4 & 40 \\ 
      9 & 138.4 & 2265 & 226.5 & 45.3 & 40 \\ 
      10 & 133.8 & 2179 & 217.9 & 43.58 & 40 \\ 
      11 & 135.09 & 2921 & 292.1 & 58.42 & 60 \\ 
      12 & 133.28 & 2531 & 253.1 & 50.62 & 50 \\ 
      13 & 114.36 & 1378 & 137.8 & 27.56 & 30 \\ 
      14 & 107.04 & 2061 & 206.1 & 41.22 & 40 \\ 
      15 & 101.99 & 1822 & 182.2 & 36.44 & 35 \\ 
      16 & 90.34 & 1941 & 194.1 & 38.82 & 40 \\ 
      17 & 83.26 & 2449 & 244.9 & 48.98 & 50 \\ 
      18 & 80.37 & 984 & 98.4 & 19.68 & 20 \\ 
      19 & 58.62 & 2491 & 249.1 & 49.82 & 50 \\ 
      20 & 64.44 & 1358 & 135.8 & 27.16 & 30 \\ 
      21 & 46.71 & 777 & 77.7 & 15.54 & 15 \\ 
      22 & 50.82 & 1187 & 118.7 & 23.74 & 25 \\ 
      X & 156.04 & 2186 & 218.6 & 43.72 & 45 \\ 
      Y & 57.23 & 579 & 57.9 & 11.58 & 12 \\ 
      mitochondria & 0.02 & 37 & 3.7 & 0.74 & 0 \\ 
      \hline
       \textbf{Total} & 3088.32 & 54083 &  ~ & ~ & 1067 \\       
      \hline
    \end{tabular}
  \end{adjustbox}
  \caption{Number of simulated number of loci per chromosome.}
  \label{Tab:Genome} 
\end{table*}

Let us first precisely define what ``genome'' means. In biology, the genome is referred to as the complete set of genetic elements of a 
given organism -- here humans. In all known cellular organisms, the genetic material is composed of deoxyribonucleic acid (DNA), which 
is the shelf of the genetic information. This macromolecule, see Fig.~\ref{Fig:Genome}~(A), is composed of two strands coiled together in 
a well-known double helix structure. Each strand is made of chains of building blocks called deoxyribonucleotides (that, for convenience, 
we shall term ``nucleotides'', even if this is not chemically rigorous), whose succession constitutes the strand sequence. Nucleotides 
(A, T, G, C) found on one strand associate with complementary nucleotides on the opposite strand (A with T, G with C and reciprocally) 
to form so-called base pairs (bp) that contribute to the double helix's stability. Because of this complementarity, complete knowledge 
of one strand sequence also provides complete knowledge of the second\footnote{This property is used during the DNA replication process 
that occurs before cell division: put simply, upon dissociation of the two strands, enzymes of the replication machinery ensures that a 
complementary strand be synthesized for each parent strand, which results in the production of two novel identical double helices containing
the same information as that initially found in the parent double helix. The two double-stranded DNA molecules can ultimately be distributed
between the two daughter cells that are therefore genetically identical.}. This will facilitate our simulations, since only the ``virtual sequence'' 
of one strand shall be considered. DNA carries genetic elements (informational sequences) that can be genes -- i. e. sequences that contain 
the instructions to synthesize proteins or non-coding ribonucleic acids --, but also regulatory sequences -- i. e. that 
participate in the control and modulation of gene expression in response to internal and/or external stimuli --, or other types of elements. 
The human genome is composed of 3 $\times$ 10$^9$~bp (two base-paired complementary strands of 3 $\times$ 10$^9$ nucleotides each) and typically 
encodes $\sim$ 22\,000 protein genes and approximatively the same number of non-coding RNA (ribonucleic acids) genes, see Tab.~\ref{Tab:Genome}. 
Genetic elements are found at defined positions along the DNA molecule and, to simplify, we shall consider that these are discrete and 
non-overlapping, even though reality is more complex. The degree of complexity as to model these genetic elements to mirror the human genome 
is far beyond the scope of this study and well beyond computing possibilities.

A straightforward method to simulate a simplified human genome is to use matrices: let A$_1$, A$_2$, ... A$_n$ be the set of individual 
positions along the DNA molecule ``A''. Those positions represent discrete and bounded blocks that are independent of each other in the 
sense that we can distinguish them by a property (their sequence, for example). As in genetics, a ``bounded block'' (of sequence) shall 
be termed a locus (plural \textit{loci}). Loci can in principle either be considered nucleotide sequence intervals of any length, meaning
that they can represent and/or contain any genetic feature or combination of genetic elements that are present on DNA. Their position 
on the chromosome is indexed in the order of their sequence by the letter $i$. We thus can create a matrix of one column and $n$ rows,
where each line represents a particular locus. For A we thus have [A$_i$]=(A$_1$, A$_2$, A$_3$, A$_4$ ... A$_n$), as shown in 
Fig.~\ref{Fig:Genome}~(B).

\subsubsection{Modeling the diploid genome and chromosomes}
\label{Improvements:chromosomes:Chromosomes}

Humans are diploids, meaning that each human cell actually carries two genomes. One copy (a so-called haploid genome of $\sim$ 3 $\times$ 10$^9$~bp) 
is of maternal origin, while the second haploid genome ($\sim$  3 $\times$ 10$^9$~bp) is of paternal origin. An individual's diploid genome 
($\sim$ 6 $\times$ 10$^9$~bp) is the result of the combination of both. As shown in Fig.~\ref{Fig:Genome}~(C), the human diploid genome is 
in fact separated into 46 physically independent DNA segments (double helices) that are called chromosomes: 23 of paternal origin and 23 of 
maternal origin. Among them, 22 paternal chromosomes and their 22 maternal counterparts are homologous, meaning that they are almost identical 
in terms of sequence, and conceptually grouped into 22 pairs of homologous chromosomes (autosomes). As an illustration, maternal chromosome 1 
has its homologous paternal chromosome 1 counterpart -- they share highly related sequence features -- and so on for chromosomes 2 to 22. The 
two remaining are sexual chromosomes (gonosomes). In humans, females carry two homologous X sexual chromosomes (XX), while males possess one X
and one small male-specific Y chromosome (XY) that are non-homologous (they differ in term of sequence, length, architecture, genetic elements 
content, etc.). Each chromosome carries its own set of genetic elements arranged along the sequence. Therefore, each chromosome can be modeled 
with a matrix of the same form as described above: for the first chromosome ``A'' we have [A$_i$]=(A$_1$, A$_2$, A$_3$, A$_4$ ... A$_n$). 
For the second chromosome ``B'', we have [B$_i$]=(B$_1$, B$_2$, B$_3$, B$_4$ ... B$_n$) and so forth. Since each chromosome has one homologue, 
it implies that genetic elements (loci, sequence features, etc…) are always found in two copies within cells (one of maternal the other of 
paternal origin), with the exception of males, for which X- and Y-specific genes are present in single copies. Chromosome A thus
has its A' homologue, represented with matrix [A'$_i$]=(A'$_1$, A'$_2$, A'$_3$, A'$_4$ ... A'$_n$), B has its homologous B' chromosome represented 
with matrix [B'$_i$]=(B'$_1$, B'$_2$, B'$_3$, B'$_4$ ... B'$_n$), etc. The complete human genome of a single individual can thus be numerically
modeled using 46 individual matrices that we store in a single C++ vector, i.e. the diploid genome.

\subsubsection{Genetic variations and alleles}
\label{Improvements:chromosomes:Alleles}

The genetic information of human individuals is never rigorously identical (they are not clones). Genetic variations exist between individuals 
and between human populations that originate from past mutations that, by descent, were transmitted to the offspring. One consequence is that in 
one individual, the two haploid genomes inherited from his/her parents are not identical. If one considers a given locus 
A$_i$ at a given position $i$ in a haploid genome (on chromosome A, for instance), it can have a given form (sequence), but another sequence 
(carrying variations of any type in various proportions) in another haploid genome (on the homologous chromosome A'). The two 
loci (A$_i$ and A'$_i$) are the same genetic information, but with different states that are termed allelic forms\footnote{Strictly 
speaking, ``homologous sequences'' stands for ``sequences that derived (through mutations) from an ancestral sequence''. Alleles refer to homologous 
sequences that are encoded at the same locus (position) of a given genome or chromosome, but that present sequence variations. 
Identical positions and similar sequences is sufficient to refer genetic elements as allelic forms.}. The two loci are termed alleles (or allelic 
forms) to one another. A given locus can have one or multiple possible allelic forms within a population. Let A$_{11}$, A$_{12}$, ..., A$_{1m}$ 
be the $m$ allelic forms that the first locus of A can take. The matrix [A$_{ij}$] therefore represents all the possible variants of 
all the loci found along A within the population, see Fig.~\ref{Fig:Genome}~(D). The set of possible alleles of all the loci of chromosome A is thus 
an $n$ $\times$ $m$ matrix, with $n$ the number of loci (blocks) along A and $m$ the number of allelic forms of a given locus. The same 
can be applied to the homologous chromosome A'. A haploid genome can thus be considered a combination of specific alleles in a given order, that we 
shall term a haplotype. Each individual is diploid, and carries two haploid genomes -- and, strictly speaking, two haplotypes --, a combination that 
is named a genotype (combination of alleles in a diploid individual). 

\subsubsection{The virtual human genome}
\label{Improvements:chromosomes:Virtual}

As stated above, the human haploid genome is composed of 3 $\times$ 10$^9$~bp and contain $\sim$ 22\,000 protein genes and an almost 
equivalent number of non-protein genes. Due to the memory-space limitations of modern computers, it is challenging to allocate several 
thousands\footnote{Typically, a 600 years-long HERITAGE run using an initial crew of 500 persons and a ship capacity of 1000 inhabitants
simulates more than 8000 individuals over 25 generations in total.} of vectors that each contains 46 $\times$ 2 $\times$ 22\,000 $\times$ $m$ 
integer values if we would consider 22\,000 blocks (corresponding to protein genes) on each haploid genome. The task would be even 
more challenging if individual nucleotides had to be taken into account (6 $\times$ 10$^9$). We thus reasoned to approximate the human
genome with a scaled-down model in order to keep the computing time reasonable. We therefore arbitrarily separated the sequences of each 
individual chromosome into N discrete blocks (where N corresponds to the number of genes of each chromosome divided by 50, see 
Tab.~\ref{Tab:Genome}, fifth column), so the number of blocks became downscaled to 2110 for the entire diploid genome, with 100 loci for 
the largest chromosome (chromosome 1). Therefore, the human haploid genome of 3 $\times$ 10$^9$~bp is, in our model, constituted of 1055 
sequence blocks that, for convenience, we also termed loci. In this way, each locus/block can alternatively be considered a single gene, 
a set of genes, or any given DNA sequence of any size with specific and defined characteristics, depending on the scale to be considered. 
We thus included in the \textit{Human} C++ class of HERITAGE (the blueprint of each numerical individual) a vector of 2110 integer 
values\footnote{We exclude the mitochondria from our calculations (see Tab.~\ref{Tab:Genome}, last column).} that is representative of 
the human genome (accounting for the homologous chromosomes\footnote{Chromosomes that belong to a single pair; all genomes are identical 
in size and have identical loci/blocks (same architecture and same organization). Their alleles can, however, be different.}). This vector 
will be filled upon the creation of the crew member, either at the beginning of the simulation or during the interstellar travel when 
reproduction will happen. The genome of each individual is stored by the program so that statistical and biological tests can be performed
during and after the completion of the simulation. The typical memory-size of one human genome stored on the computer is 4.2~ko (34.4~kb). 
Note that the 1055 loci are distributed in a given order onto 23 chromosomes and that this architecture never changes. In reality, it is not 
strictly the case, but for simplicity we imposed the architecture of genomes (order and number of loci, number of chromosomes) to remain constant.

\subsubsection{Measuring genetic diversity}
\label{Improvements:chromosomes:Measure}

A single human individual, because he/she carries two haploid genomes (he/she is diploid) independently acquired from the two 
parents, can carry two identical copies of a given locus (the same allele), or two different forms (alleles) of this locus. In the former case,
the individual is termed homozygous at this locus/position, while in the latter case, it is referred to as heterozygous at this position/locus 
(Fig.~\ref{Fig:Genome}~D). For each individual, we can therefore measure, at each locus A$_{ij}$, if he/she is heterozygous (carries two different 
alleles on chromosome A and A') or homozygous (carries two identical alleles on A and A'). From this, we can measure the individual heterozygosity 
I$_{k}$ of the k$^{th}$ individual that is the fraction of pairs of homologous loci (A$_{ij}$ and A'$_{ij}$) that are heterozygous. In the case of 
inbreeding, I$_{k}$ is expected to decrease, because two closely related individuals (that share strong similarities in terms of allelic combinations) 
tend to produce descendants that are highly homozygous (reduced heterozygosity). In this sense, I$_{k}$ is a measure of the genetic diversity at the 
individuals' scale that will be used to evaluate inbreeding, consanguinity or similar phenomena that could arise from population genetics in individuals. 

In addition, individual loci (A$_{ij}$) can have one or more allelic forms within a population. If the locus under investigation has more than 
one allelic form in the population, it is termed polymorphic. The degree of polymorphism (P) represents the fraction of loci (among the total N loci) 
that are polymorphic at the population scale. The allelic diversity (number of possible alleles) and the frequency of each allelic form within the
whole population also have to be taken into account, because they both influence the proportion of possible heterozygous or homozygous individuals 
at various positions of the genome.

Finally, the heterozygosity index (H$_{i}$) measures the proportion of individuals that are heterozygous at position $i$
on locus A$_{i}$. Since the proportion of heterozygous individuals (at a given position $i$) depends on the actual allelic diversity (number and 
frequency of allelic forms at position $i$), it is also a measure of the genetic (allelic) diversity in the population. For $m$ allelic forms of a 
given locus, there are $m$ possible homozygous and $m$($m$-1)/2 heterozygous pairs that can exist in individuals. H$_{i}$ depends on the number of
possible alleles and their respective frequencies in the population. H$_{i}$ is maximal (allelic diversity is maximal) when all allelic forms at 
position $i$ are equifrequent, with H$_{i,max,m}$=1-(1/$m$) at locus A$_{i}$. As indicated before, inbred or consanguineous populations tend to 
produce individuals that possess, on average, more homozygous positions than non-inbred populations, meaning that H$_{i}$ is expected to decrease 
at discrete positions of the genome in the case of inbred populations, a phenomenon known as the Reduction in Heterozygosity (ROH) that Smith used,
for one single locus, to evaluate the MVP of an interstellar journey \cite{Smith2014}. In HERITAGE, we can now map H$_{i}$ at all loci along the 
genome (except for sexual chromosomes) to visualize genome-scale changes in the genetic diversity of the population upon interstellar travels.

\subsection{Building the initial population}
\label{Improvements:population_ini}

The selection of the zeroth-generation for multi-generational space travels is of prime importance. First of all, one must realize 
that neither the initial crew members, nor most of the forthcoming generations, would reach the spaceship's final destination. It means 
that they would be born, raised, live, have children, and die within the limited and enclosed environment offered by the vessel without 
any possibility for leaving this protective shell or tread upon the surface of a planet, hospitable for human life or not, before arrival. 
Long-duration off-Earth space missions within the Solar System (to the Moon or Mars) are already expected to cause strong emotional, 
psychological and psycho-pathological effects due to isolation and confinement but also to inter-personal, organizational and cultural 
aspects \cite{Palinkas2001,Collins2003,Kanas2009,Landon2017,Tafforin2017}. Such a series of constraints would undoubtedly be even stronger 
and more profound for people traveling beyond the Solar System, simply because interstellar travel implies to cross unthinkable distances.
Spaceship system failure, exposure to on-board pathogens, radiation, social conflicts, external accidents, etc., would drive people, 
agencies or governments in charge of interstellar space exploration to select initial crew members with mental and psychological abilities 
that could best-fit such long-term constraints. Moreover, remoteness might favor the rise of a novel space culture with its own sociological, 
political, cultural, ethical -- and possibly linguistic -- properties and references \cite{Billings2006,Pass2007,Smith2012,Smith2014,Smith2019}, 
which would preclude any a priori (and unattainable) attempt to ``socially engineer'' an initial crew on the very long term.

Multi-generational space travel also raises biological issues regarding genetic diversity and health. In our case, ``genetic diversity'' shall 
refer, as we stated before, to the allelic diversity within the entire population enclosed in the vessel. It is described by the 
degree of polymorphism (P), the heterozygosity index of individuals (I$_{k}$) and the locus heterozygosity index (proportion of heterozygous 
individuals) at each locus (H$_{i}$). A ``genetically diverse'' population is ideally polymorphic, with a significant proportion of loci with 
multiple allelic forms that ensure that H$_{i}$ does not approaches 0 (a case arising when only one allele exists in the population at position $i$), 
implying that heterozygous positions can exist, and that I$_{k}$ remains high (individuals have multiple heterozygous positions). Note that, if P 
is high, a high proportion of all loci are polymorphic (have two or more allelic forms), which enables individuals to be heterozygous at various
positions (increased I$_{k}$), and increases the chances that a given locus be heterozygous at the population scale (measured with H$_{i}$, 
the proportion of individuals that are heterozygous at position $i$).

Why should polymorphism and heterozygosity not become too low? We already indicated that inbreeding and consanguinity, that both reduce allelic
diversity and, consequently, heterozygosity (both I$_{k}$ and H$_{i}$), have well-documented consequences on health \cite{Fareed2017} and fertility 
\cite{Charlesworth2009}. This comes from the fact that, when genetically alike individuals reproduce, they produce descendants with genomes that 
correspond to the pooling of two genetically alike haploid genomes (see below), leading to multiple homozygous positions along the diploid genome. 
Some allelic variations (that originate from past mutations) can have deleterious manifestations (phenotypes) in individuals. The effect of such
variations depends on the zygosity: deleterious dominant mutations manifest when individuals are homozygous (two identical mutated copies of the 
genetic element are present) or heterozygous (one mutated copy of the genetic element and one copy that does not possess the same mutation), while
deleterious recessive mutations have effects only when individuals are homozygous (two mutated copies). Cystic fibrosis is an example of a recessive
genetic variation that provokes a so-called genetic disease in homozygous but not in heterozygous individuals \cite{Cutting2005}, but many others 
exist. When genetic diversity decreases, such as in the case of inbreeding and consanguinity, heterozygosity tends to decrease within the population,
with homozygous positions increasing accordingly in individuals. This also increases risks to reveal deleterious recessive genetic effects/diseases. 
All possible homozygous combinations do not necessarily occur within a natural population with a large number of individuals, and associated recessive 
phenotypes (deleterious or not) therefore never or rarely manifest (from combinatorial). Therefore, even if chosen ``genetically diverse'', the initial 
crew should, in addition, include enough individuals to avoid the next generations to be affected by inbreeding and consanguinity \cite{Marin2017,Marin2018} 
that both cause I$_{k}$ and H$_{i}$ to decrease. I$_{k}$ and H$_{i}$ can also be affected by strong stochastic variations in allele 
frequencies that could lead to random fixation (one allele becomes the only allelic form) or loss of alleles, due to a reduced number of possible mating 
combinatorial \cite{Smith2014}, a process that is referred to as ``genetic drift'' \cite{Motro1982}. Since the initial crew will necessarily be small 
(limited resources, space, etc.), this will restrict mating possibilities between individuals and potentially affect I$_{k}$ and H$_{i}$ (reduce 
heterozygosity) and lead to inbreeding and/or consanguinity. 

The initial crew would thus be regarded as a minimal viable population (MVP) \cite{Jamieson2012,Frankham2013}, in which genetic diversity (P, I$_{k}$ 
and H$_{i}$) and the number of individuals would have to be determined to reduce risks of loss of heterozygosity (decrease of I$_{k}$ and H$_{i}$) and
consequently of inbreeding and consanguinity. In order to ``stabilize'' a selected initial allelic diversity in the initial population, the number of 
individuals shall be sufficient to reach, or at least approach, the Hardy-Weinberg equilibrium, a state under which alleles frequencies remain stable 
throughout generations within the population \cite{Mayo2008}, which should also stabilize P, I$_{k}$ and H$_{i}$.
 
The selection process of initial crew members could integrate tests to choose ``genetically-healthy'' candidates with no known deleterious genetic 
variations. In comparison to psychological tests, DNA sequencing technologies, clinical and genetic tests could more easily help determine whether the 
candidate or her/his offspring carry one or several genetic markers linked to known genetic disorders. However, things are far from being that simple:
\begin{itemize}
  \item First, mutations in genes or genetic elements that generate allelic diversity can, of course, be detrimental to health, with phenotypes that 
  express as well-known hereditary/congenital diseases. These genetic variations could, in principle, be excluded from the initial population 
  to avoid highly deleterious genetic disorders. However, even if they were, \textit{de novo} spontaneous mutations could put them back into 
  the population's allelic pool during the course of the journey, especially those that are known to occur with high frequencies on Earth. 
  \item Second, mutations that are known to be associated with deleterious phenotypes can produce highly variable phenotypes (biological manifestations) 
  in various individuals, depending on their own genetic background (genotype) \cite{Zlotogora2003,Fournier2017}. If the mutation of a genetic element 
  is dominant, its associated phenotype will express even if only one of the two copies in the diploid genome is mutated; however, if it is recessive, 
  only those individuals that possess two mutated copies will express the phenotype. This implies that novel homozygous combinations, naturally arising
  in the spaceship or originating from inbreeding and/or loss of heterozygosity could reveal unanticipated and unpredictable phenotypes, including diseases 
  that, by definition, could not be detected as such when the initial population is constituted. In addition, the effect of a given dominant or recessive
  mutation not only depends on the hetero- or homozygous state of an allele, but also on the overall genetic background of individuals, that is, on other
  variations that are present across the diploid genome of an individual and that influence phenotypical manifestations. The same mutation can thus be entirely 
  neutral (no phenotypical or fitness effect), advantageous or deleterious at various degrees depending on individuals' genetic compositions. Cystic fibrosis 
  \cite{Cutting2005} is affected by such genetic influences that modify clinical outcomes and severity of the disease \cite{ONeal2018}, but this is true 
  for any phenotypical trait.
  \item Third, gene expression strongly depends on the environment (temperature, pressure, gravity, pollution, diet, quality and amount of food, radiation, 
  stress, etc.) or developmental stage of an individual. The manifestation of a phenotype associated with a given mutation therefore depends 
  on the expression pattern and timing of the mutated gene, but also on the effect that it has on the capacity of the gene's expression product (protein, RNA)
  to fulfill its function. Also, mutations in regulatory genetic elements can modify the expression pattern of one to several (sometimes hundreds of) genes 
  in response to environmental, hormonal (external) or cellular (internal) signals and lead to unpredictable phenotypes, depending on what regulatory circuit 
  and/or tissue, cell type is/are affected. Combined with the effect of the genetic background (other variations), one can understand that the effect of 
  mutations and/or combination of mutations (or alleles) is not easy -- and strictly speaking, impossible -- to predict for one individual and, moreover, 
  for an entire population \cite{Zlotogora2003,Fournier2017}. Since environmental conditions influence phenotypes, prediction of the effect of mutations
  on health, fertility or life expectancy is highly uncertain. This is also true for already existing genotypes, with known associated phenotypes (on Earth), 
  that will be placed under novel environmental conditions (spaceship) and likely for all possible novel genetic combinations (genotypes).
\end{itemize}

From those facts it is clear that it would be almost impossible to predict or anticipate the rise of novel phenotypic manifestations (deleterious, neutral or 
advantageous) on-board. That is to say, it would be merely impossible to begin with a set of starting individuals (and genomes) who are predisposed 
towards generating a so-called healthy offspring. All in all, choosing a ``good'' starting population is equivalent to choosing an MVP \cite{Jamieson2012,Frankham2013},
i.e. gathering enough individuals and allelic diversity to avoid loss of heterozygosity, inbreeding and consanguinity over time and to keep this 
diversity stable until arrival. The goal would be to favor the allelic diversity so that the genetic combinatorics repertoire of individuals remains high enough 
to provide an even higher collection of possible phenotypic manifestations under the environmental conditions of the spaceship, with the expectation that, 
among them, the fewest would be deleterious. Note that, beyond the interstellar journey, having a highly diversified population at arrival is, from the genetic 
point-of-view, also critical to establish a long-term viable colony \cite{Smith2014}, since, again, the settlers would remain separated from other human 
populations at best for long durations, but most likely forever.

Careful selection of favorable genetic characteristics of a starting population in a eugenic (ethically disputable) way -- as some have proposed -- would therefore 
be highly speculative, if not unwise, since already existing genotypes that fit Earth's conditions could randomly and unpredictably result in detrimental as well 
as neutral or advantageous expressed phenotypes on-board under non-terrestrial conditions, diets or radiation. Similarly, choosing or engineering advantageous 
genetic backgrounds to influence or drive future favorable genetic combinations (genetic engineering) would be equally unrealistic -- apart from its ethical 
disputability -- and irrelevant, given the random processes involved in the generation of the offspring (see next section and Appendix~A) that would shuffle
genotypes over time and produce novel genotypes, submitted again to unpredictable genetic interactions and random effect of the environment. 
If we add the equally random (naturally or intentionally introduced by genetic engineering) mutations that could have random effect placed in a random genotype 
of an individual living in an ever-changing and randomly varying environment, one understands that any genetics-based idealized short- or long-termed projection 
would be impossible. 

Because of the complexity inherent to the genotype/phenotype/environment relationships, at the present stage, we consider in HERITAGE that the
various allelic states and/or haplotypes and genotypes (allelic combinations in dipoids) in our code have neutral effects. This means that the 
combinations of alleles within genomes do not lead to genetic disorders neither in initial crew members that carry them, nor in 
their offspring, where those combinations change. In other words, there is no negative (deleterious) or positive (advantageous) selection of 
alleles, haplotypes or genotypes over time as a result of environmental, genetic, developmental, physiological, etc. constraints. 
Such hypothesis, very often used in population genetics simulations, will be examined in the second paper of this series. 

To build the initial population, we first define a standard reference human genotype by setting all alleles to 0. We will use it for comparison 
with the $n^{\rm th}$ generation for the purpose of detecting variation and changes in the genetic structure/composition of the population. Then, 
in order to construct the carefully hand-picked, initial population, we decided to let the user select one of two options. 
\begin{itemize}
  \item The first option consists in a starting population in which each initial crew member has a completely randomized genotype 
  (combination of alleles at the diploid state). In this population, individuals carry on average 5\% differences (variations) with
  respect to the standard reference human genotype \cite{Redon2006}. To do so, we randomly assign an allelic state comprised between 
  1 and 9 to randomly picked loci along all the chromosomes. This allows us to build genomes with realistic amounts of variations but with the 
  drawback that there is no genetic history behind the various crew members. This means that we do not expect recognizable allele patterns
  between crew members that account for the existence of genetic lineages at the beginning of the simulation. Although this is likely 
  to be an idealized population, it will help us check the validity of our code in Sect.~\ref{Improvements:meiosis}.
  \item The second option is meant to construct a ``non-random'' zeroth-generation population with a chosen amount of 
  variations with respect to the standard reference human genotype (in which all loci are set to 0). For example, a variation level of 
  20\% means that individuals carry, on average\footnote{In natural 
  human populations, two individuals can carry millions of genetic differences at the nucleotide level (base pair differences) that, in 
  comparison to the size of the genome (3 $\times$ 10$^9$~bp) represent around 0.2\% differences \cite{Auton2015}. In our case, remind 
  that we separated chromosomes (of size $S$, in bp) into $N$ discrete blocks, where $N$ arbitrarily corresponds to the number of genes 
  ($p$) of each chromosome divided by 50 ($N$=$p$/50) (see Tab.~\ref{Tab:Genome}, fifth column), meaning that each locus of chromosome~1 
  (100 loci) contains approximatively 2.5 million bp (length $L$ corresponds to $L$=$S$/$N$=50$S$/$p$, where $S$ is the chromosome's size 
  in bp, $p$ the number of genes). Five millions of bp changes (0.2\% differences) between two individuals, if evenly distributed over the 
  1060 loci of the haploid genome, would represent hundreds of thousands of bp changes in chromosome~1 alone, implying thousands of differences 
  in one single locus between two individuals. With such an amount of differences between two alleles, then, the number of possible allelic 
  states for one locus becomes really high. We restricted those differences to only 10 possible allelic states (with no information on the 
  actual amount of differences between them) for simplicity. When we produce a population in which 0.5\% of loci can carry allelic variations, 
  one therefore understands that, in HERITAGE, it actually represents far less variation between individuals than in real populations, making 
  them genetically very closely related. For this reason, we also permitted to produce populations in which variation can be selected up to 80\% 
  (a variation level that, even with multiple allelic states for each locus, remains well below the actual variation that exists in nature).}, 20\% of loci that adopt allelic states different from 0.
  Of course, the less variation, the closest individuals should be considered from the genetic point of view. An allelic variation of 0.5\%, 
  for example, simulates a population constituted of individuals that share close genetic ancestry, i.e. a ``low diversity'' population. 
  With increasing variation levels, populations mimic more diverse groups, in which close genetic ancestry between individuals becomes 
  less probable. For each population type, we first created pools of 100 individual genotypes with a variation level of x\%. We then 
  crossed these 100 genotypes in a randomized fashion: either 2, 3 or 4 sub-genotypes are mixed to simulate successive generations of 
  tribes/populations mating, resulting in five final populations. All these genotypes and populations are stored and used as a different 
  starting material each time we run HERITAGE. To constitute the initial crew, we randomly choose $k$ individuals among these 5 reference 
  populations. This makes it possible to account for the fact that these populations, even if they are the initial ones, are themselves 
  the result of a complex (and \textit{common}) genetic history, with varying levels of genetic relatedness.
\end{itemize}

\begin{figure*}[!t]
  \begin{subfigure}[t]{.7\textwidth}
    \centering
    \includegraphics[trim = 0mm 12mm 0mm 12mm, clip, width=12cm]{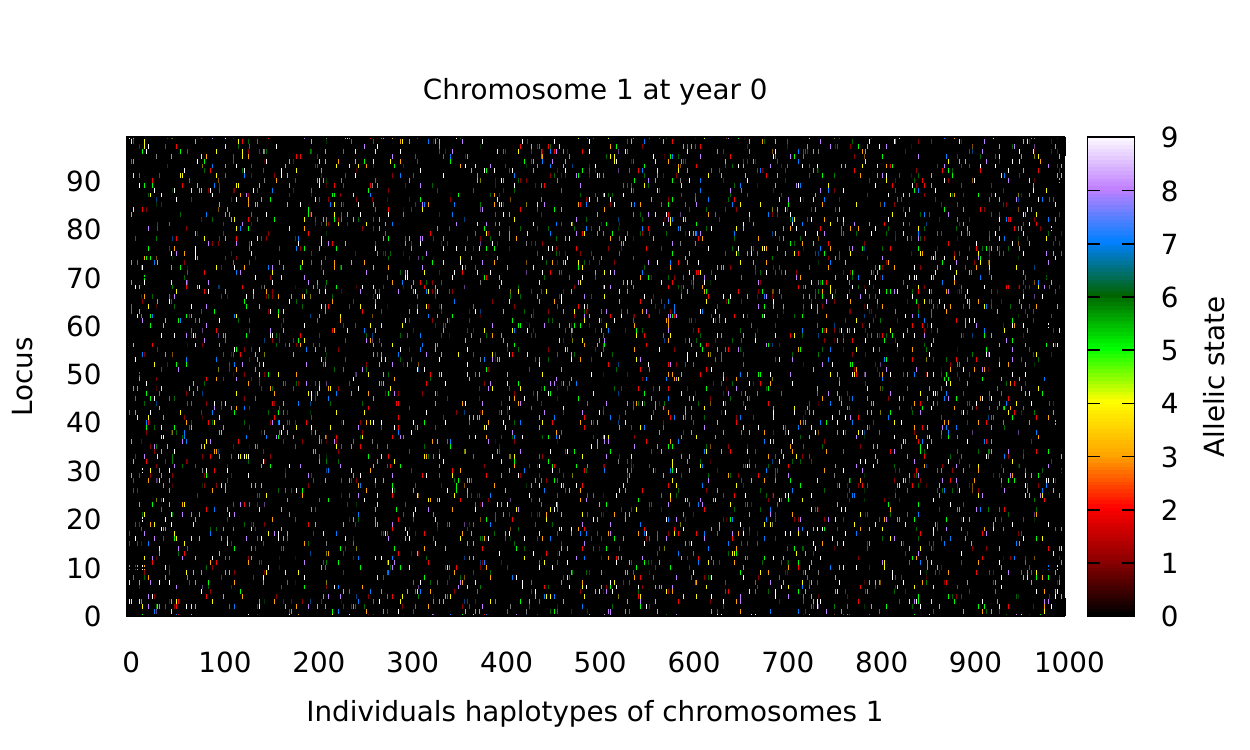}
  \end{subfigure}
  \hfill
  \begin{subfigure}[t]{.3\textwidth}
    \centering
    \includegraphics[trim = 12mm 0mm 30mm 0mm, clip, height=2.2cm, angle=90]{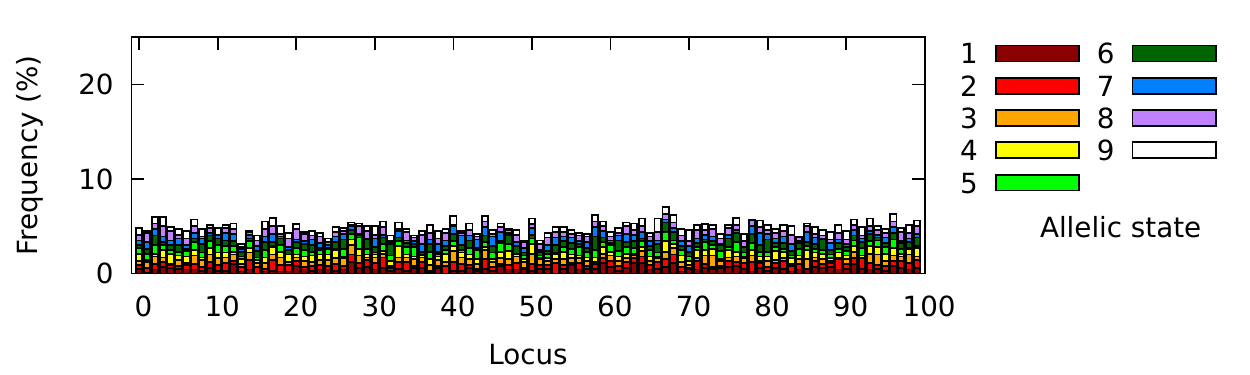}
  \end{subfigure}

  \medskip

  \begin{subfigure}[t]{.7\textwidth}
    \centering
    \includegraphics[trim = 0mm 0mm 0mm 12mm, clip, width=12cm]{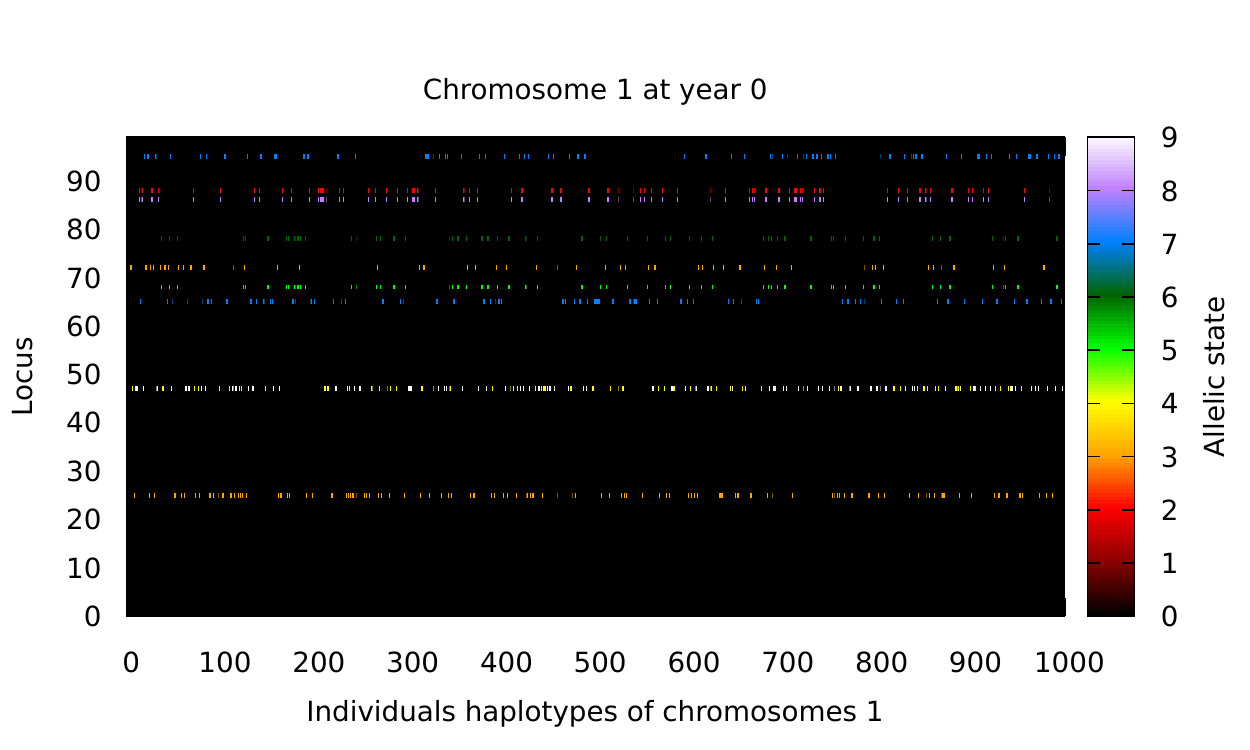}
  \end{subfigure}
  \hfill
  \begin{subfigure}[t]{.3\textwidth}
    \centering
    \vspace{-6.1cm}\includegraphics[trim = 0mm 0mm 30mm 0mm, clip, height=2.2cm, angle=90]{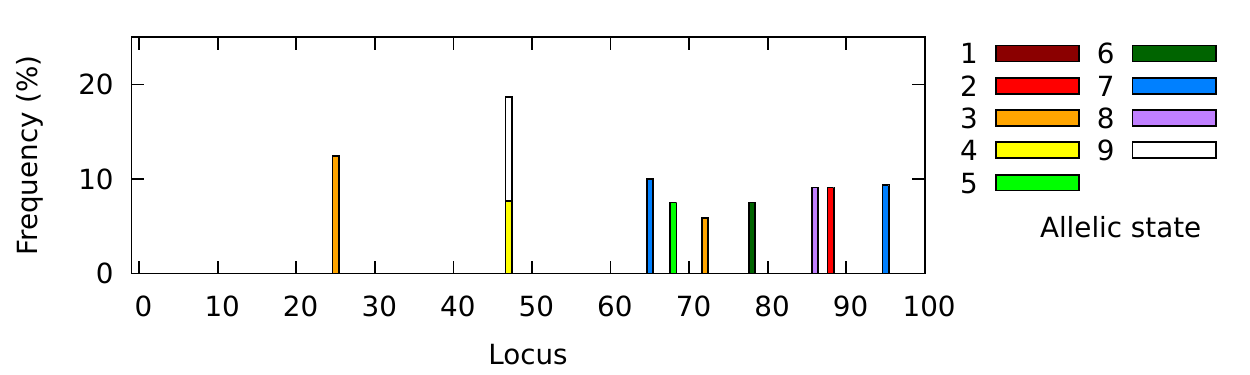}
  \end{subfigure}
  \caption{Haplotype heat maps of all chromosomes 1 found in an initial population of 250 women and 250 men. It presents 1000 haplotypes 
  that correspond to the 1000 chromosomes 1 of the 500 diploid (initial) crew members. For each locus, a color code indicates its allelic state. 
  The top figure shows the randomized population, in which 5\% of variations (randomly distributed) are found within each genome, relative 
  to the standard reference human genotype (all loci set to 0). The bottom figure shows a population for which individuals already share 
  genetic patterns and whose genomes show an allelic variation of only 0.5\% with respect to the standard reference human genotype (low 
  diversity population). A stacked histogram on the right of each heat map allows to better visualize the distribution of the allelic forms
  for which the allelic state is non-zero (black alleles are not displayed for simplicity). Each bar represents the frequency of each allele 
  with the same color code as in the heat map.}
  \label{Fig:Initial_pop}%
\end{figure*}

We programmed HERITAGE to automatically generate heat maps and stacked histograms representing the allelic composition (haplotype) of each 
chromosome, both at the beginning and at the end of the mission, together with graphs showing the degree of polymorphism P of 
the population, the heterozygosity index of individuals (I$_{k}$) and the heterozygosity index for each locus along chromosomes (H$_{i}$). 
In the rest of this publication, we will only show the results for chromosome 1 (for haplotypes heat maps and  
stacked histograms) for space saving purposes but all the chromosomes data are simultaneously plotted by the code. 

In Fig.~\ref{Fig:Initial_pop}, we show on a heat map the 1000 different allelic patterns (haplotypes) of chromosomes 1 of an initial (zeroth) 
population of 500 individuals (250 males, 250 females). Since each of the 500 individuals is diploid (and has 2 chromosomes 1), 1000 chromosomes 1
are displayed. Allelic states found in the modeled loci are represented with a color code. We present the case of a randomized initial population 
(top figure, 5\% of all loci carry variations) and the case of a non-random initial crew with much less allelic diversity (bottom figure, 
0.5\% variations). Both constitute extreme test cases that shall help present the possibilities offered by the improvements of the code
to visualize changes in the allelic composition of traveling populations. Examples of non-random populations with variation levels of 
5, 20 and 50\% and pre-existing allelic patterns are also provided in Appendix~B. In all cases, the initial population (500 individuals) is larger 
than the MVP thought to be needed for interstellar travel (100 individuals), such as determined in \cite{Marin2018} to match more 
``classic'' populations \cite{Hardy1908,Weinberg1908,Fisher1923,Wright1931}.
 
In the case of a population with a randomized allelic diversity set to 5\%, i.e. without previous genetic history or designed patterning, we see 
in the haplotypes heat maps and stacked histograms of Fig.~\ref{Fig:Initial_pop} that most of the loci found on chromosome 1 have multiple allelic 
forms (polymorphism P is high), each with low and similar frequencies (within statistical fluctuations), which is characteristic of the random 
attribution of allelic states to loci. In the case of a low diversity population (allelic diversity set to 0.5\%), the haplotypes heat map and 
stacked histogram show that allelic patterns do exist at the population level, which originates from pre-existing allelic patterns (haplotypes) 
implemented in ancestral populations. Also, only few allelic forms (1 to 3) in only a few loci (10 out of 100 on chromosome 1 in 
this example) exist (low polymorphism). Populations with variation levels of 5, 20 and 5080\% were also tested (see Appendix~B); as expected, 
polymorphism increases with variation levels, as well as the heterozygosity index of each locus (H$_{i}$), that reflects the increased allelic 
diversity, which translates into an increasing heterozygosity index at each locus (H$_{i}$) that describes the proportion of heterozygous individuals
at those positions.

\subsection{Gamete production, meiosis and formation of the n+1 generation}
\label{Improvements:meiosis}

\begin{figure*}[!t]
\centering
  \includegraphics[trim = 0mm 0mm 0mm 0mm, clip, width=\textwidth]{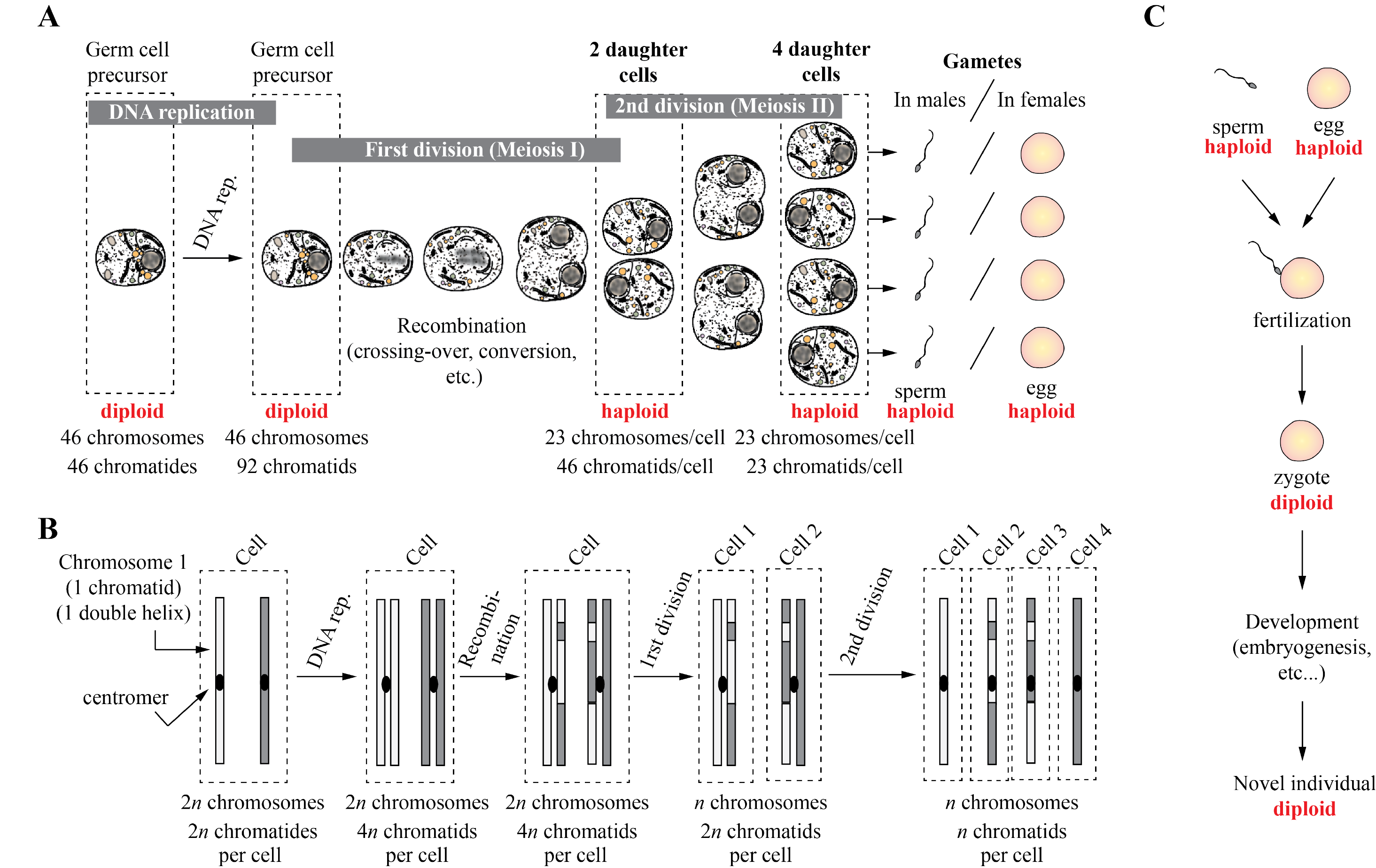}
  \caption{Principles of meiosis (formation of haploid egg and sperm cells from diploid precursor germ cells, 
	   panel A) and of genetic recombination followed by chromosomes and chromatids random shuffling 
	   (panel B). Panel~C shows the formation a diploid individual from the haploid egg and sperm cells.}
  \label{Fig:Meiosis}
\end{figure*}  

Once our initial population is created, we can run HERITAGE to generate the n+1 generation. A complete description of HERITAGE can be found in 
\cite{Marin2017,Marin2018,Marin2019,Marin2020}, so we shall simply summarize the main steps for offspring's generation. The code randomly selects
two humans (one female and one male), checks that they are alive and within their procreation window, and determines by random draws if the two 
successfully mate. The code accounts for all necessary age-dependent biological parameters such as fertility, chances of pregnancy, miscarriage 
rate, etc. and checks whether the offspring is not inbred (within the security margins imposed by the user, using Wright's genealogical parameters
\cite{Wright1922}). The new crew member is assigned an identification number. Various anthropometric parameters (weight, height, basal metabolic 
rate, etc.) are computed together with the life expectancy of the individual. Before the upgrades presented in this paper, we randomly assigned 
the sex of the offspring and did not account for her/his genetic heritage.

Now that each crew member of the zeroth-generation has a specific genotype, we can follow the rules of heredity to properly create the genotype of
the offspring. The first step is to produce gametes (ova/eggs and spermatozoa/sperm, since only the variations present in these sexual cells are 
transmitted to the offspring). Gametes are produced from so-called germ line precursor cells, the only cells that can undergo meiosis. Meiosis is 
the process of double-cell division that allows switching from a diploid cell (two homologues for each chromosome) to four haploid cells (with a 
single chromosome of each kind in each cell, see \cite{Alberts2002,Baudat2013,Bolcun2018} and Fig.~\ref{Fig:Meiosis} A). We recall that each human 
cell has 46 chromosomes that correspond to 22 (males) or 23 (females) homologous pairs. Chromosome 1 therefore exists in two homologous forms 
1a and 1b, one (1b) from the father, the other (1a) from the mother. Their sequences are homologous, which means that they are similar but may 
carry sequence (state) differences, that is, they carry different haplotypes. It is the same for chromosomes 1 to 22 (1a, 1b, 2a, 2b, 3a, 3b ..., 22a, 22b). 
This is different in the case of the sex chromosomes because female organisms have two homologous X chromosomes (Xa and Xb, from mother and father 
respectively), while males have one X chromosome (from the mother) and one Y chromosome (from the father) that are not homologous to each other. 

During meiosis, the germ cells (cells that give rise to the gametes of an organism that reproduces sexually) start by duplicating all
present DNA. This means that the 46 chromosomes present in the cell become duplicated. Chromosome 1a will therefore be duplicated in 1a and 1a', 
the homologous chromosome 1b in 1b and 1b', etc. Each chromosome therefore now possesses two chromatids (a and a', b and b', two DNA helices, clones 
of each other, see Fig.~\ref{Fig:Meiosis} B) which remain connected to each other by what is called the centromere\footnote{This is the reason 
chromosomes are usually drawn as elongated Xs, with each side being a chromatid and the cross being the centromere, where both duplicated DNA 
molecules remain bound.}. So, at this point, the amount of DNA is doubled, as it is the case for any cell division. Once everything is doubled, 
each pair of homologous chromosomes gets closer and both homologues undergo what is called \textbf{homologous recombination} (crossing-over event). 
In fact, one of the chromatids of one homologue interacts with one of the chromatids of the other homologue to form pairs of chromatids. There are 
four possible combinations: 1a with 1b, 1a with 1b' or else 1a' with 1b, or 1a' with 1b' (interactions of 1a with 1a' or 1b with 1b',
although possible, are neglected, since they do not produce changes in allelic patterns/haplotypes). Only one of the combinations is chosen at random 
and the same is true for other chromosomes. These interactions occur over a certain lengths $l$ (the same on both chromatids). Homologous recombinations
occurs within this interval. This means that the DNA sequences at these interaction zones make it possible to exchange the sequences contained in the 
interval of length $l$ between the two chromatids. For example, for the interaction of 1a with 1b, the exchange of sequences over a length $l$ between 
these two chromatids causes the passage of a segment from 1a to 1b, and reciprocally from 1b to 1a. It is the same for the other three combinations, 
if they are chosen. If the sequence exchange is unidirectional, that is, a sequence of length $l$ of chromatid 1a is shifted to chromatid 1b and replaces
the original sequence, but reverse direction does not occur (1a remains unchanged), it is a phenomenon called \textbf{conversion} \cite{Baudat2013,Bolcun2018}. 
By and large, the sequence contained in the interval $l$ of chromosome 1a imposes the sequence that will be present in 1b, but not the other way round. 
Overall, note that for any starting genotype constituted of two independent haplotypes, the homologous recombination and conversion (exchanges between 
homologous sequences) that takes place in germ cells will change the combination of alleles (haplotypes) found on individual chromosomes 
and randomly create genetic diversity, i.e. novel alleles combinations along chromosomes.

Once recombination and/or conversion are done, the homologous chromosomes are randomly separated and distributed in two different daughter cells. 
Therefore, each of the two daughter cells will contain 23 chromosomes with 2 chromatids each. The choice between 1a and 1b, 2a and 2b, etc. is 
entirely random, which, again, creates diversity. After this distribution, a second division of meiosis takes place for each of the two daughter 
cells. During this division, in each of the cells, the chromatids of each chromosome are separated and distributed in two daughter cells randomly. 
We thus obtain, from one starting germ cell, four daughter cells in total, each having 23 chromatids from the 46 starting chromosomes.
These cells are haploid because they contain only one chromosome of each species and no longer two, as in the beginning. The process is the same for 
egg formation and sperm formation, so we do these genetic tasks for both the mother and the father (see Fig.~\ref{Fig:Meiosis} C). Homologous 
recombination, conversion, random separation of homologous chromosomes and random separation of chromatids shuffles the pre-existing genetic 
information, i.e. modify haplotypes of the final sexual cells (sperm or egg).

We must highlight the fact that the mechanism of meiosis, leading to four genetically different gametes, occurs for a single starting germ cell 
but there are thousands of germ cells, and millions of random possibilities of genetic shuffling in each one of them, so the combinatorics is 
really gigantic. This is why we use the full power of the Monte Carlo method to test all possible events and have a representative outcome of 
the meiosis. In HERITAGE, before mating, the code now uploads the vectors containing the mother's and father's genomes and creates haploid female 
(ovum/egg) and male (spermatozoon/sperm) gametes throughout the process described above. The algorithm performs recombination between pairs of 
homologous [X$_{ij}$] and [X'$_{ij}$] over intervals of length $l$ that are randomly selected between 3 $\le l \le$ 8 loci at the same time 
according to a discrete uniform distribution along the chromosome to make sure they do not always occur at the same place. In the code, there 
are 1 to 5 exchange areas per homologous pairs and the number of trades is also chosen at random. The code also allows conversion, i.e. the
unidirectional exchange, over small areas (1 to 2 loci at maximum) with a known frequency of $\sim$ 10$^{-7}$ \cite{Hogstrand1994,Stahl2001,Harpak2017}, 
so about 7.18 times over the entire genome in our simplified model of the human genome. For mating (and creation of a new individual), two final
gametes, after meiosis, meet and pool their two haploid genomes to form a diploid genome containing two homologous chromosomes of each type. This
genome is stored in a new vector of 2110 integers and is saved under the identification number of the child. The genome of the offspring is thus 
a novel combination of those haploid genomes from the two gametes, themselves selected from random but biologically-realistic processes. 
Pooling two Xs or one X and one Y makes it possible to determine the sex of the offspring in a sensible way, without imposing a ratio that, 
biologically speaking, does not actually exist since biological sex is due to this random pooling. Using this scheme, each novel individual 
resulting from the pooling of two haploid genomes from its parents' sexual cells contains a novel and unique genotype, that is the result of 
the combination between two unique haplotypes obtained through meiosis in the parents. 

\begin{figure*}[!t]
  \begin{subfigure}[t]{.7\textwidth}
    \centering
    \includegraphics[trim = 0mm 12mm 0mm 12mm, clip, width=12cm]{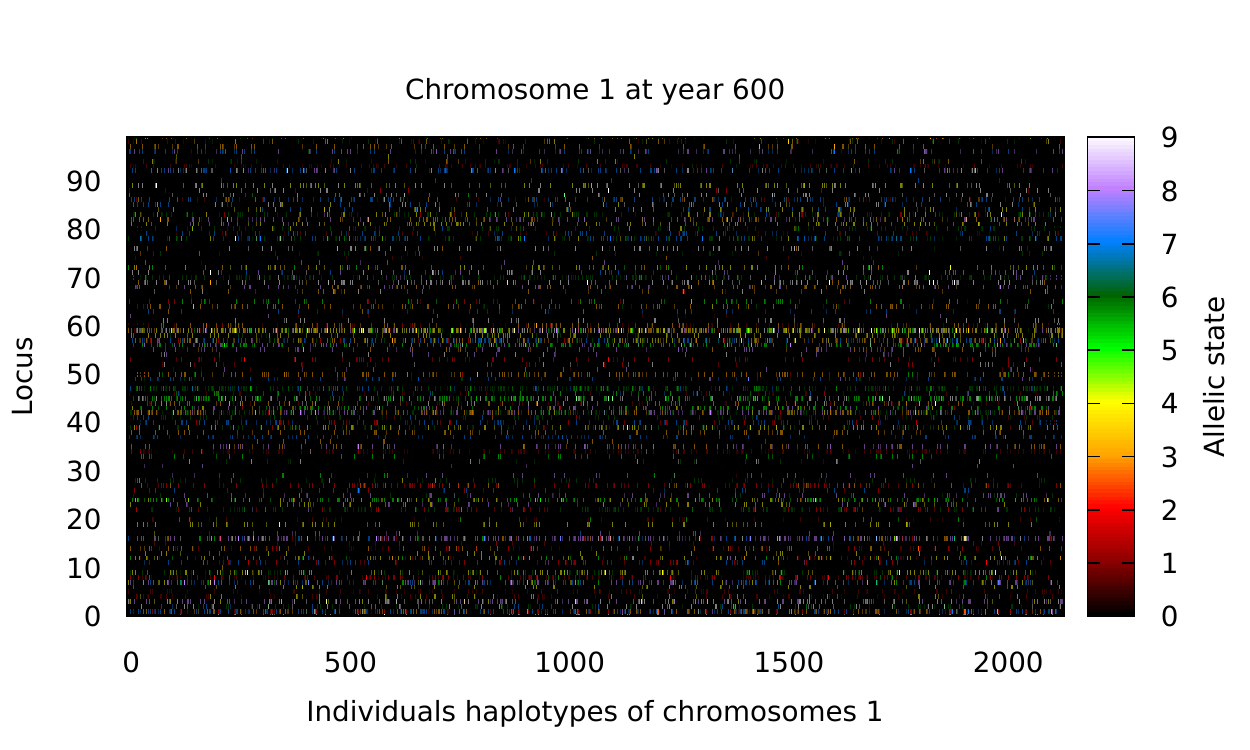}
  \end{subfigure}
  \hfill
  \begin{subfigure}[t]{.3\textwidth}
    \centering
    \includegraphics[trim = 12mm 0mm 30mm 0mm, clip, height=2.2cm, angle=90]{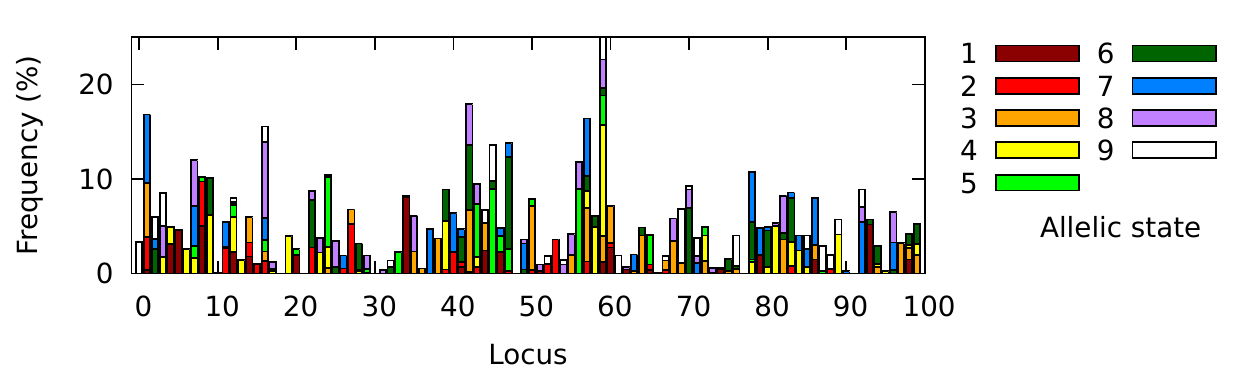}
  \end{subfigure}

  \medskip

  \begin{subfigure}[t]{.7\textwidth}
    \centering
    \includegraphics[trim = 0mm 0mm 0mm 12mm, clip, width=12cm]{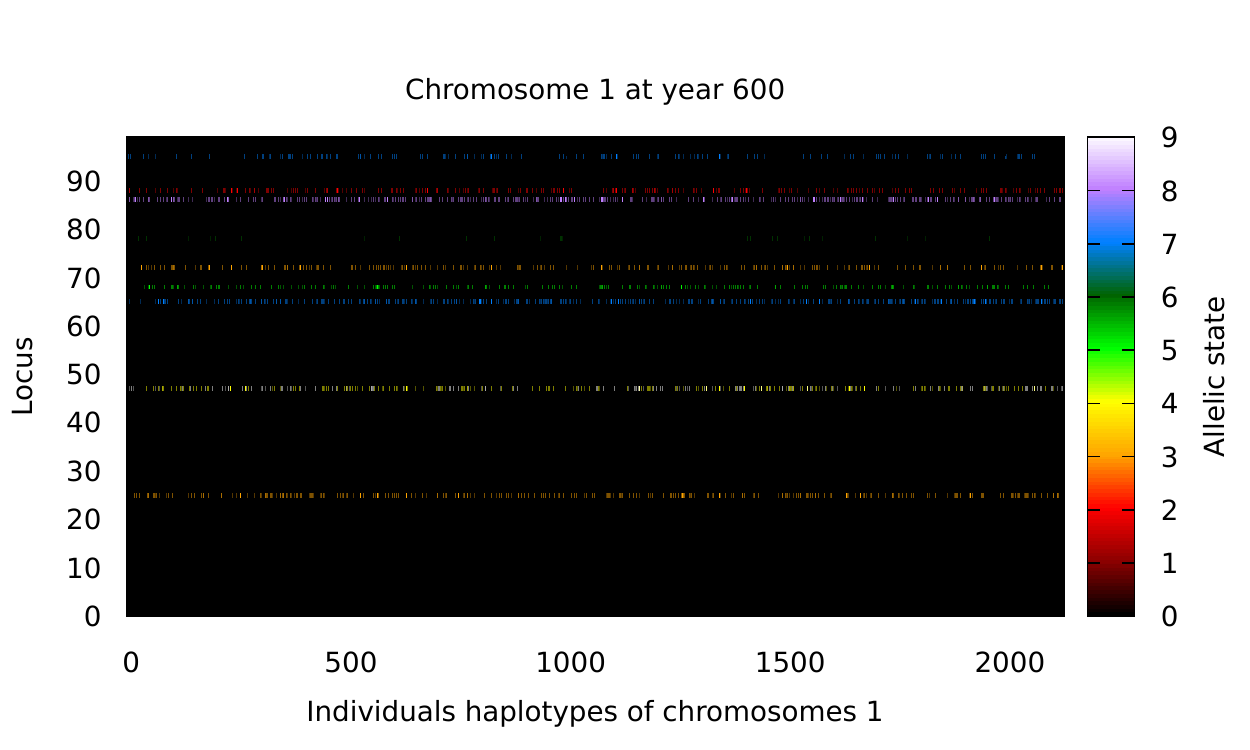}
  \end{subfigure}
  \hfill
  \begin{subfigure}[t]{.3\textwidth}
    \centering
    \vspace{-6.1cm}\includegraphics[trim = 0mm 0mm 30mm 0mm, clip, height=2.2cm, angle=90]{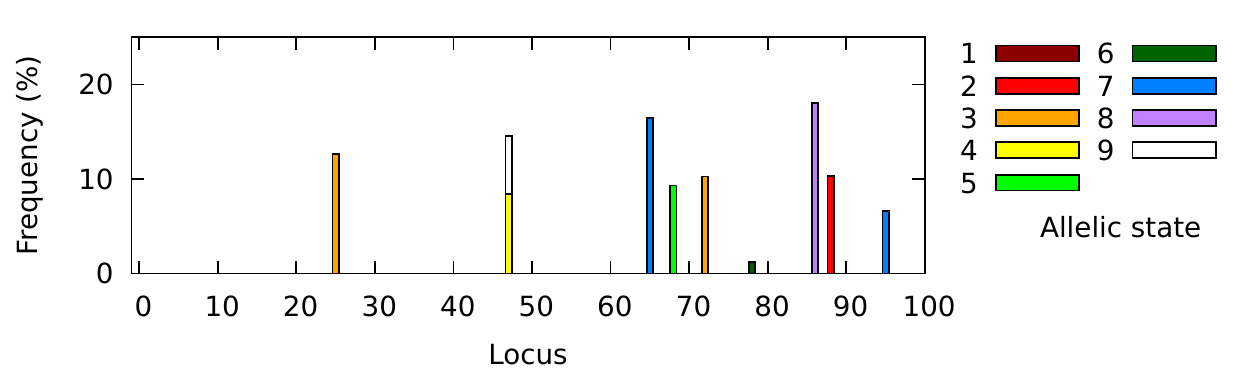}
  \end{subfigure}
  \caption{Haplotypes heat maps of all chromosomes 1 in a final population of approximately 1080 persons after 600 years of space travel 
  under little-to-no cosmic ray radiation (no mutational effects). The top panel shows the genetic composition of a final population that descends from an initial 
randomized population in which the starting allelic diversity was 5\%, with no pre-existing genetic patterns. Results (to be compared with Fig.~\ref{Fig:Initial_pop}, top panel)
show that this final population, that is the result of a 600 year-long and complex genealogical history, now presents recognizable allelic patterns that are visible on the heat map and
highlighted by significant changes in the number and frequencies of alleles for discrete loci. The bottom panel shows the genetic composition of a final population that descends from an initial 
low diversity population in which the starting allelic diversity was 0.5\%, with pre-existing genetic patterns. Results (to be compared with Fig.~\ref{Fig:Initial_pop}, bottom panel)
show that allelic patterns did not change significantly, but that allelic frequencies changed.}
  \label{Fig:Final_pop}%
\end{figure*}

Fig.~\ref{Fig:Final_pop} presents the results of 600 years of breeding for the enclosed population in the spaceship, following the newly implemented 
biological laws. The ship's volume capacity was fixed to 1200 inhabitants at maximum, with a security threshold of 90\% to avoid overpopulation. 
Consanguinity was not allowed (up to first cousins once removed or half-first cousins, i.e., a consanguinity factor of 3.125\% or below) 
in this simulation. The procreation window was selected to be between 30 and 40 years old according to the results from our previous publications 
\cite{Marin2018,Marin2019}. The top panel shows the genetic composition of a final population that descends from an initial randomized population 
in which the starting allelic diversity was 5\%, with no pre-existing genetic patterns (random assignment of allelic states for ~5\% of loci). 
This final population is the result of a 600 year-long and complex genealogical history that produced novel genotypes through meiosis (genetic recombination, 
chromosomes and chromatids random shuffling) and random mating. The results (to be compared with Fig.~\ref{Fig:Initial_pop}, top panel) show that, contrary 
to the initial population, recognizable allelic patterns are now visible on the heat map of the final population. This is also highlighted by significant 
changes in the number and frequencies of alleles of discrete loci. Several alleles have been favored -- others eliminated -- by crossing-over, conversion 
and mating histories and the global genetic diversity of the final population shows clear differences with respect to the completely randomized distribution 
from year 0. While our theoretical population is not realistic (no pre-existing patterns), the results highlight the fact that the biological laws we have 
implemented work well and can generate novel allelic patterns that are the result of genetic recombination and shuffling mechanisms (see also Appendix~A). 
The genetic diversity of the final population is still close to the initial value of 5\% since neomutations were not permitted. This denotes that the number of starting 
individuals was enough, and that this number remained enough to stabilize allelic diversity, as if the population approached the Hardy-Weinberg equilibrium.
The bottom heat map and stacked histogram show the genetic composition of a final population that descends from an initial low diversity population in which 
the starting allelic diversity was 0.5\%, with pre-existing genetic patterns. Results (to be compared with Fig.~\ref{Fig:Initial_pop}, bottom panel) highlight 
that allelic patterns did not change significantly, but that allelic frequencies did. This comes from the fact that the starting allelic patterns and allelic 
diversity were highly reduced, which decreased the combinatorics possibilities, contrary to the above-mentioned randomized population. However, we observe 
stochastic changes in allele frequencies: some alleles are much less present than in the beginning of the journey, while some others have increased. This 
illustrates the genetic drift that resulted in changes in allelic frequencies from sampling effects (mating, recombination, etc.) in small populations. 
Genetic drift promotes intergroup differentiation in the long term. Here, the final genetic diversity is close to the initial one due to the absence of 
spontaneous or cosmic-ray induced mutations.

\subsection{Introducing mutations}
\label{Improvements:mutations}

The stability of the genetic information is central to the normal function of cells and, more importantly to the reproduction of living organisms. It is
therefore of vital importance to maintain genomic stability within the somatic (non-sexual) cells that constitute all the organs, tissues and structures 
of the body. Genomic stability avoid deleterious alterations of homeostasis and proliferation that could cause, among others, carcinogenesis, but also 
in sexual (germ line) cells that ensure transmission of this information to the offspring. This stability is ensured in both types of cells by the enzymatic 
(protein) machineries that replicate DNA \cite{Ganai2016} and by those that correct inevitable replication errors (mutations \cite{Baarends2001,Chatterjee2017}), 
a balance that results in very low mutation rates \cite{Campbell2013}. Mutations are modifications of DNA sequences. They naturally and continuously occur 
within cells as a result of physicochemical constraints imposed to DNA itself but also to the cellular machineries and processes that ensure DNA replication and 
transmission \cite{Maki2002}. When they arise in germ cells, they are the cause of changes in the genetic composition of the offspring that are responsible
for the emergence of novel polymorphisms (sequence variants), i.e. alleles, that make individuals genetically different (in addition to the genetic 
recombination and shuffling due to meiosis). 

When a cell divides (mitosis), it must duplicate its entire genome and DNA polymerases are the enzymes (proteins) that catalyze the synthesis (polymerization)
of a novel DNA strand using a single stranded DNA template, free nucleotides and the ``complementarity rules'' \cite{Ganai2016}. They possess ``proofreading''
activities that ensure correction of several types of errors, such as mismatches (wrong base-pairings), during or after replication \cite{Ganai2016,Bebenek2018}. 
Oxidative, chemical or radiation-induced stress can alter the nucleotides chemistry, thereby influencing their base-pairing properties and leading to mismatches 
that lead to so-called punctual mutations. These stresses can also provoke various types of covalent cross-links between nucleotides and/or strands, as well as 
DNA single or double strand breaks that all alter the integrity of the genetic information and perturbs faithful DNA replication. This can lead to small or 
large sequence deletions (losses), insertions or duplications \cite{Okayasu2012,Sankaranarayanan2013,Rak2015}. Those novel mutations (neo-mutations) may have 
deleterious effects \cite{Ku2013}. Sometimes, alterations lead to large scale chromosomal rearrangements, i.e. changes in the architecture of chromosomes (whole 
region duplications, deletions, translocation of sequence elements from one chromosome to the other, fusions of chromosomes, etc.) that can all affect the overall 
physiology or even survival of cells \cite{Weckselblatt2015}. Moreover, chemicals- or radiation-induced stress can dramatically increase abnormal chromosome 
and/or chromatids segregation during mitosis (cell division) or meiosis (germ cells' specialized division, see above), leading to erroneous partition of the 
genetic material and of chromosomes, inducing aneuploidies (wrong number of chromosomes in one cell) that are highly detrimental to somatic cells \cite{Newman2019} 
or to reproduction when they occur in sexual cells \cite{Taylor2014,Gunes2015}. 

However, cells are equipped with DNA repair proteins that detect and resolve mismatches and other types of DNA alteration, such as nucleotides chemical alterations, 
strands breaks, etc. triggered by oxidative-, chemical- or radiation-induced stresses \cite{Maki2002,Okayasu2012,Sankaranarayanan2013,Rak2015,Chatterjee2017}. 
Overall, DNA polymerase proofreading activities and DNA repair processes keep mutation rates very low, in the order of 1 erroneous nucleotide incorporated every
10$^{8}$ to 10$^{10}$ added nucleotide during replication \cite{Baarends2001,Ganai2016,Bebenek2018}. As a result, the mutation rate is on the order of 10$^{-8}$ 
single nucleotide mutation per base pair per generation in germ line cells in humans (depending on the age of the individual), while small deletions or insertions 
are on the order of $\sim$ 10$^{-9}$. Duplication or deletion of regions of 50~bp or more occur at rates of $\sim$ 10$^{-4}$ -- 10$^{-2}$, depending on the sequence's 
length \cite{Campbell2013}. Despite those frequencies, punctual mutations (single nucleotide variants) and small deletions/insertions are, by far, the most frequent, 
likely because large scale changes (deletions, insertions or displacement of larger sequences) are deleterious and cannot be transmitted to or by the offspring \cite{Campbell2013}. 

Let, again, ``A'' be the first chromosome. We have [A$_i$]=(A$_1$, A$_2$, A$_3$, A$_4$ ... A$_n$) the set of loci indexed in the order of localization along A.
Let [A$_ij$] be the chromosome A containing $n$ loci for which each can take the state $j$, that varies from 0, the ``reference state'', to $m$, with, in our 
case $m$ taking one single value on one haplotype, comprised between 0 and 9 (only 10 different alleles of a given locus are authorized to exist in the initial 
population). If a mutation occurs in germ line cells inside one of these loci, then its state $j$ takes an integer value $k$ that, by definition, has to be different
from the other pre-existing values (for other allelic forms). In reality, a mutation could, in principle, change the allelic state of one allele to a state that 
is identical to another, already existing allele, within the population. However, given the size (in~bp) associated to each locus, such mutations are highly improbable
and shall be neglected. We take a mutation rate (that is also the mutation probability) of 1.2 $\times$ 10$^{-8}$ single nucleotide change per base pair per generation 
\cite{Campbell2013}. For simplicity, all \textit{de novo} mutations shall account for punctual mutations or small deletions/insertions. Larger DNA rearrangements, 
such as large scale deletions/insertions or even gene duplications, chromosomal changes (translocations, fusions, etc.) will not be considered in this work

\subsection{Impact of cosmic rays}
\label{Improvements:cosmic}

In the interstellar medium, a continuous flux of atomic nuclei and high energy (relativistic) particles have been detected \cite{Linsley1961,Linsley1963}. 
These cosmic radiation consist mainly of charged particles \cite{Compton1934,Mewaldt1994,Auger2014}: protons (88\%), helium nuclei (9\%), antiprotons, electrons,
positrons and neutral particles (gamma rays, neutrinos and neutrons). The sources of the most energetic radiation (whose energy exceeds 10$^{20}$~eV) are not
yet fully identified but are likely to be extra-galactic in nature (either from active galactic nuclei or the collapse of super-massive stars \cite{Auger2017}).
These particles are extremely harmful to Earth-like life \cite{Nelson2016} because they carry enough energy to ionize or remove electrons from atoms, possibly leading
to DNA breaks and/or alterations \cite{Radiation1990}.

Radiation (heavy ions, ionizing radiations, etc.) can induce chemical group modifications on nucleotides and change their base-paring properties, but also
produce cross-links between nucleotides, and/or induce DNA single- or double-strand beaks \cite{Frankenberg1990,Okayasu2012,Sankaranarayanan2013,Rak2015}. These 
alterations may be repaired by the DNA repair machineries \cite{Sankaranarayanan2013,Chatterjee2017}. However, space conditions such as microgravity and/or radiation 
can cause DNA damage and affect DNA repair mechanisms to the extent that genetic mutations may accumulate over time, especially in somatic (non-sexual) cells 
\cite{Moreno2017}. Somatic cells in the human body (or any embarked animal or plant) would be the main victims of such radiation, with effects that depend on
the type of radiations and localization and properties of affected tissues. DNA alterations caused by space radiation are not necessarily repaired \cite{Moreno2017}
by the dedicated cellular machineries \cite{Okayasu2012,Rak2015}, which can perturb DNA replication and cause genomic instability \cite{Tang2014}, thereby 
leading to mutations of various possible types in somatic cells that may produce detrimental effects such as cellular deregulations, cell death and cancers 
(carcinogenesis) \cite{Durante2008,Barcellos2015,Sridharan2015,Sridharan2016,Li2018}. These may also be the cause of various health issues and pathologies 
\cite{Tang2017,Tang2018,Squillaro2018}, including nervous system alterations \cite{Cekanaviciute2018,Jandial2018}. Such somatic cell DNA alterations would not
be transmitted to the offspring. If occurring in exposed embryos or fetuses \cite{Jacquet2004,Brent2013}, they could also trigger developmental deregulations, 
malformations or cancers.

Of course, if genetic alterations caused by space radiation occur in germ line (sexual) cells, they are, only in this case -- if not repaired --,
transmitted to the offspring \cite{Nakamura2013}, potentially leading to congenital diseases and/or other abnormalities \cite{Tang2017,Tang2018,Mishra2019}, 
if the associated mutations are not neutral. Studies on mouse models show that mutation rates increase in germ line cells when they are chronically exposed to 
ionizing radiations, especially in males, and that the effect becomes much greater for acute exposures and the same can likely be extended to humans 
\cite{Nakamura2013,Adewoye2015}. Fortunately some of the space radiation and high energy particles are deflected by the solar wind and, at ground level on 
Earth, they are widely dispersed by the magnetosphere or blocked by the atmosphere and its particles in suspension. Because of this, cosmic radiation only 
accounts for 13 to 15\% of terrestrial radioactivity \cite{Delves2007}. However, in space, the annual flux of cosmic radiation received by astronauts is 
greater and therefore represents a danger \cite{Nelson2016}. This danger is all the greater as one moves away from the Sun and its natural protection. It is 
therefore understandable that cosmic radiation (and its impact on the human genome and overall health) is a considerable risk for any interstellar travel.

\begin{figure*}[!t]
  \begin{subfigure}[t]{\textwidth}
    \centering
    \includegraphics[width=0.45\linewidth]{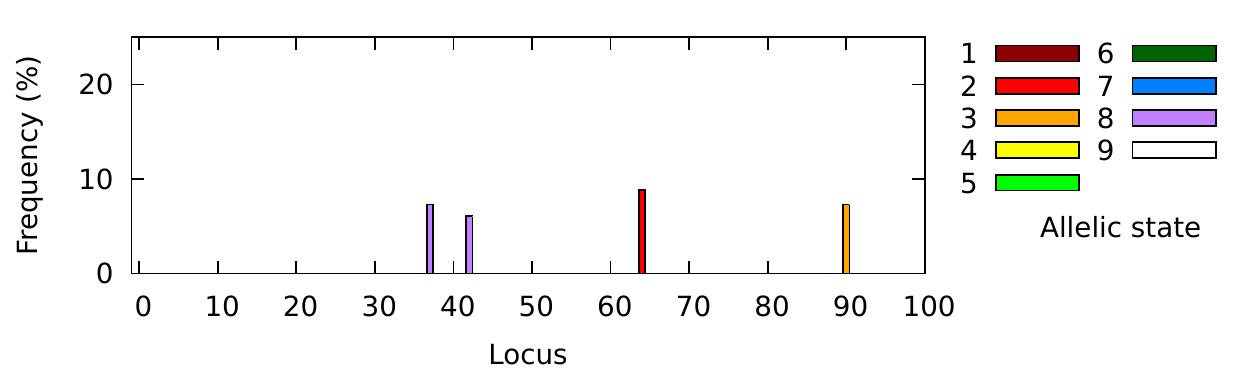}
    \includegraphics[width=0.45\linewidth]{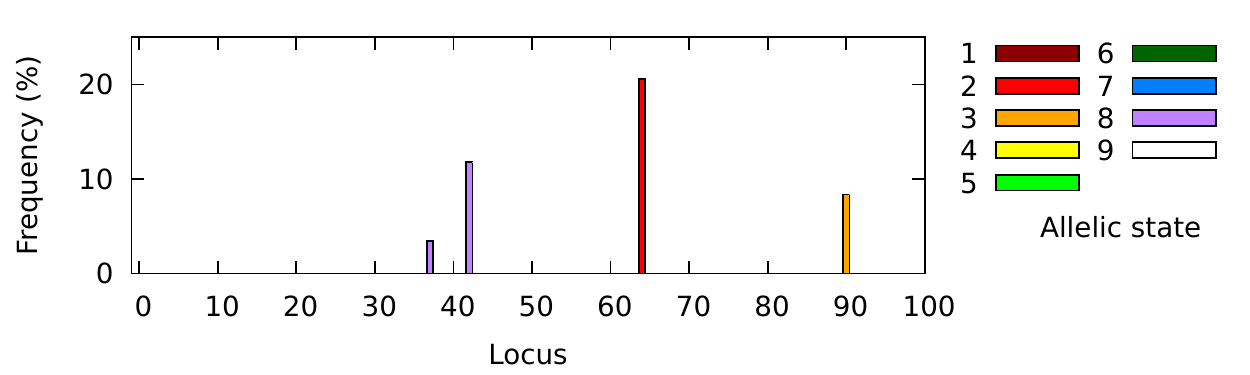}
    \caption{Constant annual radiation dose: 0.3~mSv.}
  \end{subfigure}

  \medskip

  \begin{subfigure}[t]{\textwidth}
    \centering
    \includegraphics[width=0.45\linewidth]{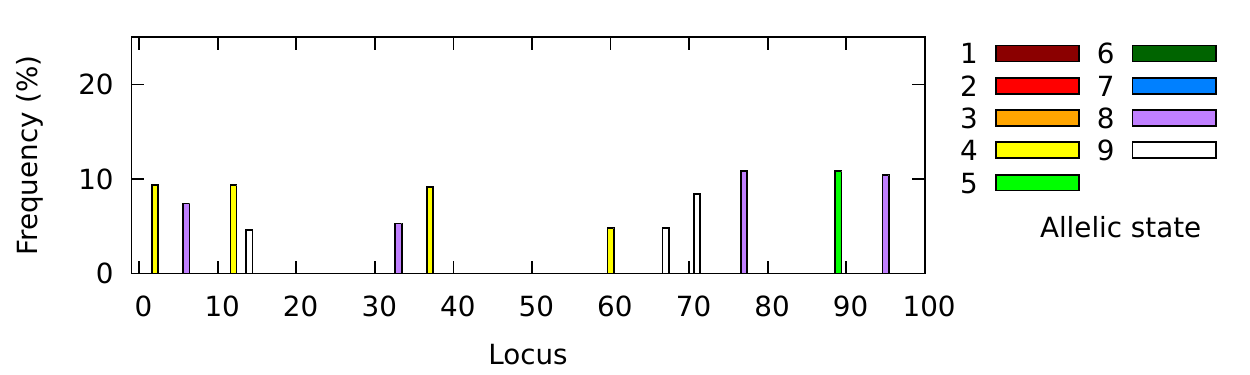}
    \includegraphics[width=0.45\linewidth]{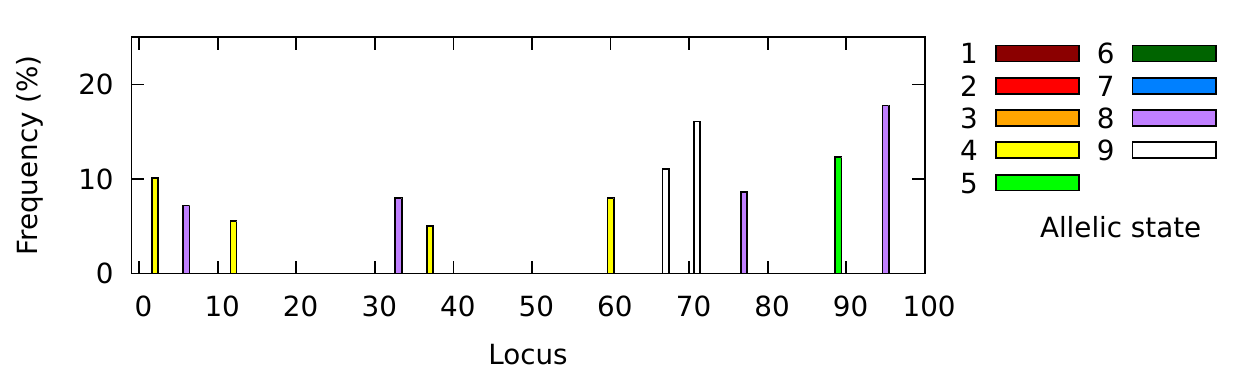}
    \caption{Constant annual radiation dose: 3~mSv.}
  \end{subfigure}

  \medskip

  \begin{subfigure}[t]{\textwidth}
    \centering
    \includegraphics[width=0.45\linewidth]{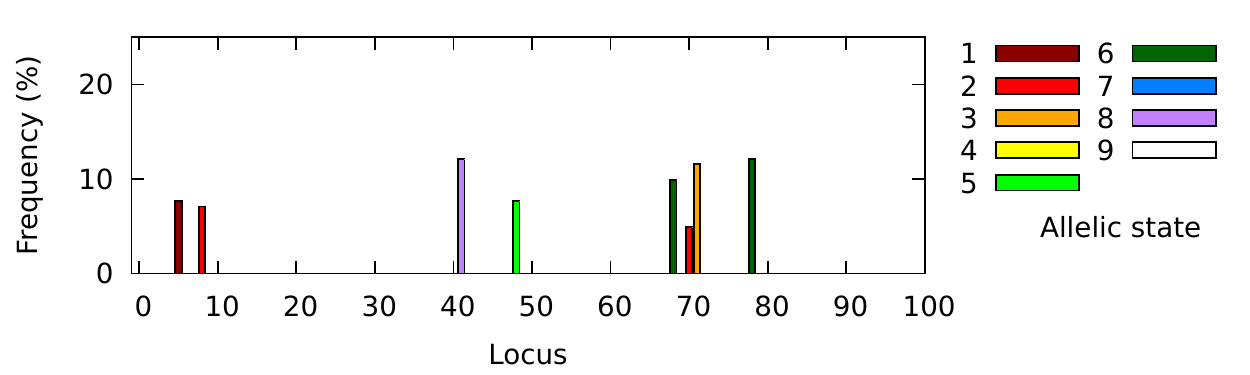}
    \includegraphics[width=0.45\linewidth]{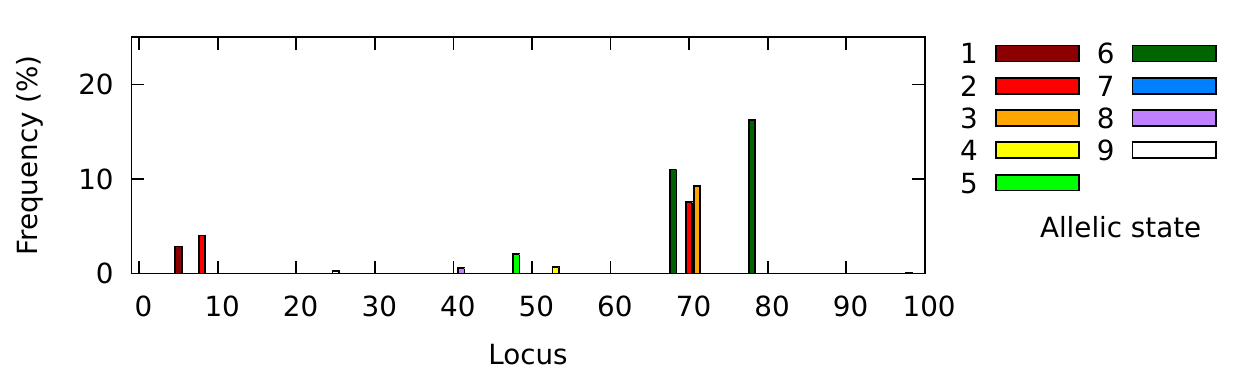}
    \caption{Constant annual radiation dose: 30~mSv.}
  \end{subfigure}  

  \medskip

  \begin{subfigure}[t]{\textwidth}
    \centering
    \includegraphics[width=0.45\linewidth]{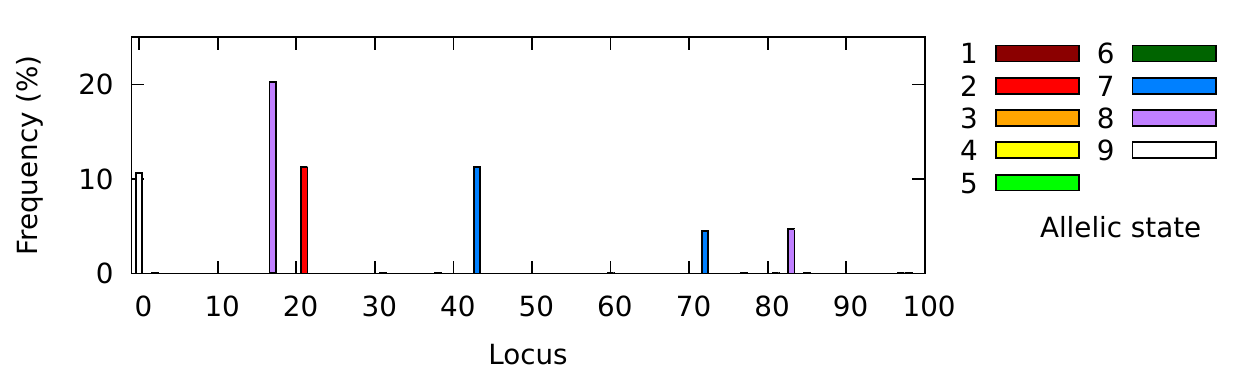}
    \includegraphics[width=0.45\linewidth]{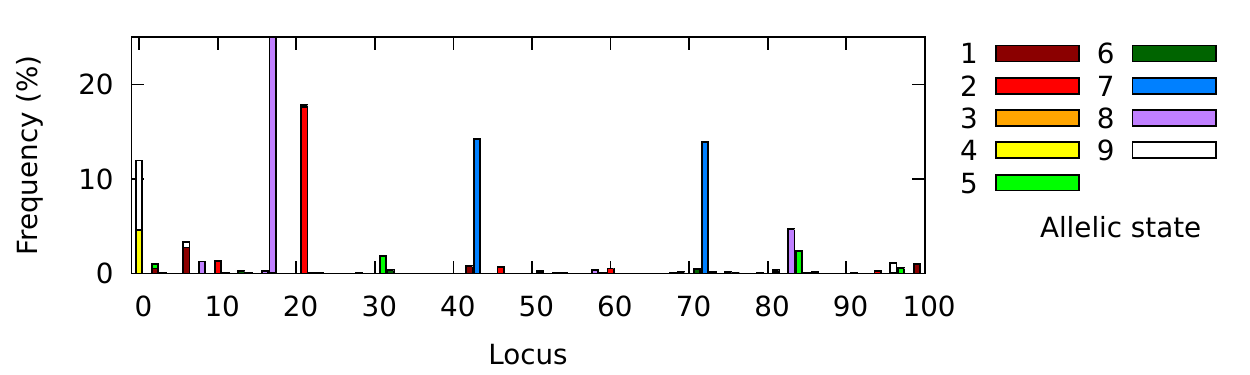}
    \caption{Constant annual radiation dose: 300~mSv.}
  \end{subfigure}    
  
  \caption{Effect of radiation on the overall allelic composition. Stacked histograms show the frequency 
	   of alleles found on chromosomes 1 for an initial, gender-balanced, ``low diversity'' population 
	   of 500 crew members (left figures) and for the final final populations of approximately 1080 
	   persons after 600 years of space travel (right figures). Each row corresponds to a different 
	   constant annual radiation dose: 0.3, 3, 30 and 300~mSv.}
  \label{Fig:Mutations}%
\end{figure*}

For this reason, we decided to take into account radiation-induced mutations in our recent upgrades of HERITAGE. To do so, we allow the user 
to fix an annual equivalent dose of cosmic ray radiation (in milli-Sieverts) at the beginning of the simulation. This represents the effectiveness
of cosmic ray shielding of the spacecraft. This value can be changed during the interstellar travel to simulate the degradation of the 
shielding material/technology, but also to mimic a nuclear disaster from, e.g., the propulsion system or a nearby and unexpected supernova event.
In the framework of the Earth's magnetosphere and atmospheric protection, the annual dose of radiation is of the order of 0.3 -- 4.0~mSv in
European countries \cite{Cinelli2017}. This corresponds to a mutation rate that is less or equal than 10$^{-3}$ per generation per individual
\cite{Ebisuzaki2014}, much more than the estimated ~10$^{-8}$ \cite{Campbell2013} under normal terrestrial conditions. We thus include in our 
simulation an additional random draw that is compared to the mutation rate scaled with respect to the annual dose of radiation, so that larger 
cosmic ray doses imply larger mutation rates. Note, however, that this is a simplification since the mutation rate as a function of 
cosmic ray impacts in deep space is yet to be measured and understood. The number of loci randomly affected by mutations is determined by the combination of the annual 
radiation dose and the mutation rate. In the case of 0.3~mSv per year, less than one locus is affected per generation per individual. In our model, 
any mutation of any kind that occurs within a locus $i$ that has a state $j$ shall be indicated by a change in the value of $j$, with the only 
restriction that the value must be different from those already present.

We ran HERITAGE for four different initial populations of 250 women and 250 men with a pre-existing genetic history (the ``low diversity'' 
population option). The space travel duration was set to 600 years. The ship's maximum capacity, overpopulation threshold and authorized 
consanguinity are similar to those simulated in Sect.~\ref{Improvements:meiosis}. With all the new biological upgrades, the codes now takes 3.6 
times longer to complete. A single-run simulation (no iterations of the same trip) is achieved in 22 seconds using the input parameters described 
above. In Fig.~\ref{Fig:Mutations}, we present the effects of neomutations on the human genome after 600 years of space travel with four different
constant annual radiation dose: 0.3, 3, 30 and 300~mSv. The first and second doses correspond to the annual background radiation on Earth at 
sea level and in US countries \cite{NCRP1987}. The third dose is representative of about 3 months on-board of the International Space Station 
\cite{Cucinotta2008} and the fourth dose to about 500 days on Mars \cite{Chancellor2018}. Beyond the effects of genetic drift -- that can change 
the frequency of pre-existing alleles --, we can see that mutations are very rare within 600 years for radiation doses of 0.3 and 3~mSv. The 
mutations either did not affect the genome, or were randomly lost during genetic recombination and chromosome shuffling (meiosis) or from biased 
sampling during mating. Very low-frequency neomutations emerged in the case of an annual radiation dose of 30~mSv and are still visible after 
600 years. On average, they are well below the frequencies of the alleles that were initially present in the starting population. When considering 
an annual dose of 300~mSv, neomutations become more populated, impacting the genetic composition of the final population in a more substantial 
fashion, although novel alleles remain low-frequency variations. 

Again, note that, like for allele combinations, neomutations that change the genotypes found in and transmitted by individuals are all neutral, 
with no deleterious (negative) or advantageous (positive) effects. Therefore, they do not affect the offspring (diseases, reduced life expectancy, etc.)
or the probability that descendants can reproduce (sterility, fertility, etc.). All mutations that become transmitted after genetic recombination, 
chromosome shuffling  (meiosis) and random mating therefore remain present in the population's genetic pool, unless they are randomly lost according
to the same genetic (meiosis) and reproductive (mating) mechanisms. Of note, those mutations that become transmitted to the offspring originate 
from changes in the haploid genomes of germ cells. However, we must remind that mutations also accumulate in somatic (non-sexual) cells of individuals 
during and as a function of their lifetime. This, in reality, would likely lead to cancers or other physiological perturbations, a fact that we do not 
take into account and that could also influence the transmission of germ cell-specific mutations, in addition to the effect of mutations acquired from 
ancestors. At high doses (300~mSv), individuals in the population could therefore be strongly affected by mutation-induced pathologies that affect the 
soma (e.g. cancers), i.e. somatic cells, which could change life expectancy, health, fertility, etc. in the whole population. Germ cells-specific 
mutations can be transmitted to the offspring and affect children with genetic diseases that can themselves change life expectancy, health, fertility, 
and even the capacity of cells to repair radiation-induced DNA alterations (increased mutational rate). This would most likely strongly and durably 
affect the entire population, with a time-dependent and cumulative worsening that could eventually completely wipe out the crew. It is interesting to 
note that our simulations demonstrate that at levels superior to 30~mSv, the human genome suffers numerous genetic changes (at the population and 
generation scales) that could be fatal, which is in perfect agreement with the regulatory dose limits of radiation workers (50~mSv) defined by federal 
(i.e., the Environmental Protection Agency -- EPA --, the Nuclear Regulatory Commission -- NRC -- and the Department of Energy -- DOE --) and state 
agencies (e.g., Agreement States) to limit cancer risk.

\section{Genetic effects over 600 years of interstellar travel}
\label{Model}

\subsection{Demographic results}
\label{Model:Demography}
Now that HERITAGE is able to compute, manipulate and store genetic data, we decided to run it in the context of a 600 years space travel 
towards any interesting target. For continuity purposes, we kept the same HERITAGE parametrization as before (250 women and 
250 men for the first generation, consanguinity factor below 3\%, etc.) and concentrate on the analysis of the population demographics.
We simulate a catastrophic event at year 350 that will wipe out 30\% of the population chosen at random. This will allow us to see the effect 
of a so-called ``bottleneck event'' (that affects the genetic composition of a population without selective effects on genes, i.e. rapid 
catastrophic events) in addition to genetic drift and mutations on the global genetic (allelic) composition of the final deme. We consider 
a state-of-the-art radiation shield so that the annual equivalent dose of cosmic ray radiation is similar to the Earth radioactivity background
at sea level (0.3~mSv). The initial crew is young (20 years on average), carefully picked from five different existing populations (the
``low diversity'' option) but without family connexions at this point. We use the adaptive social engineering principles established in our 
series of publications \cite{Marin2017,Marin2018}: each woman can have 3 $\pm$ 1 children over the course of her life but if overpopulation 
onsets the code will reduce this value so that there will be internal population regulation. In comparison with our previous calculations, 
we decreased the standard deviation of the female and male life expectancy (from 15 to 5) in order to better mirror current reality
\cite{INSEE2018,Ho2018}. We also extended the procreation period from 30 -- 40 years to 18 -- 40 years in order to mitigate the sibships effect 
\cite{Moore2003}. To calculate the total energy expenditure of the crew per year, we consider that the population is vigorously active 
between age 20 -- 45 and less active before and after. We will loop HERITAGE over one hundred iterations since this is enough for reasonable 
demographic estimates \cite{Deasy2000,Marin2019}. However, we must note that each iteration of the code will now produce different initial 
population genetics. This will become useful for determining the slow changes in the genetics of the crew throughout the space travel. 
Tab.~\ref{Tab:Parameters} lists all the parameters that we fixed before starting the simulation. Extensive explication, details and description 
of the parameters are given in \cite{Marin2017,Marin2018,Marin2019,Marin2020}. 

\begin{table*}[!t]
  \centering
    \begin{tabular}{lrl}
	\textbf{Parameters} & \textbf{Values} & \textbf{Units}  \\
        \hline
        Number of space voyages to simulate & 100 & -- \\
	Duration of the interstellar travel & 600 & years \\
	Colony ship capacity & 1200 & humans \\
	Overpopulation threshold & 0.9 & fraction \\	
	Inclusion of Adaptive Social Engineering Principles (0 = no, 1 = yes) & 1 & -- \\	
	Genetically realistic initial population (0 = no, 1 = yes) & 1 & -- \\
	Number of initial women & 250 & humans \\
	Number of initial men & 250 & humans \\
	Age of the initial women & 20 $\pm$ 1 & years \\
	Age of the initial men & 20 $\pm$ 1 & years \\	
	Number of children per woman & 3 $\pm$ 1 & humans \\
	Twinning rate & 0.015 & fraction \\
	Life expectancy for women & 85 $\pm$ 5 & years \\
	Life expectancy for men & 79 $\pm$ 5 & years \\
	Mean age of menopause & 45 & years \\
	Start of permitted procreation & 18 & years \\
	End of permitted procreation & 40 & years \\
	Initial consanguinity & 0 & fraction \\
	Allowed consanguinity & 0 & fraction \\
	Life reduction due to consanguinity & 0.5 & fraction \\	
	Possibility of a catastrophic event (0 = no, 1 = yes) & 1 & -- \\
	Fraction of the crew affected by the catastrophe & 0.3 & fraction \\
	Year at which the disaster will happen (year; 0 = random) & 350 & years \\
	Chaotic element of any human expedition & 0.001 & fraction \\
        \hline
    \end{tabular}
    \caption{Input parameters of the simulation. The $\mu$ $\pm$ $\sigma$ values shown for
	    certain parameters indicate that the code needs a mean ($\mu$) and a 
	    standard deviation value ($\sigma$) to sample a number from of a normal 
	    (Gaussian) distribution.}
    \label{Tab:Parameters}
\end{table*}

\begin{figure*}[!t]
  \begin{subfigure}[t]{.48\textwidth}
    \centering
    \includegraphics[width=\linewidth]{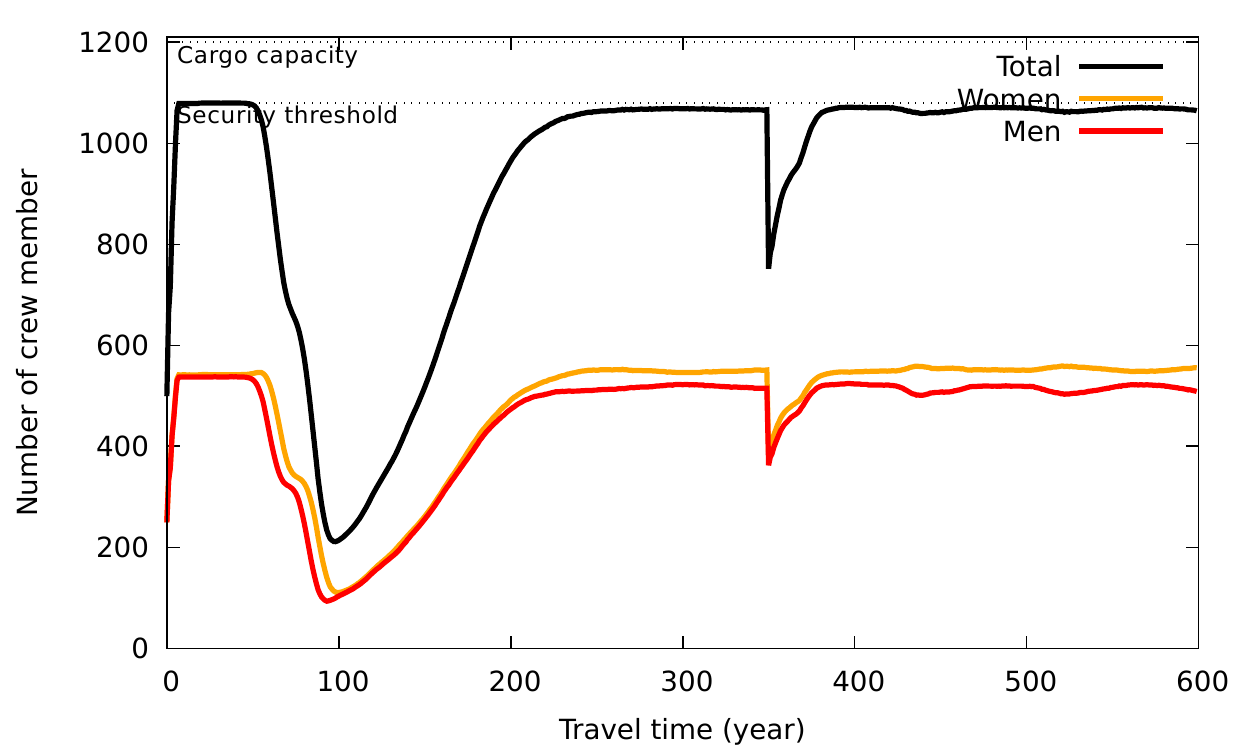}
    \caption{Crew evolution in terms of population number (orange: women, 
	     red: men, black: total).}
  \end{subfigure}
  \hfill
  \begin{subfigure}[t]{.48\textwidth}
    \centering
    \includegraphics[width=\linewidth]{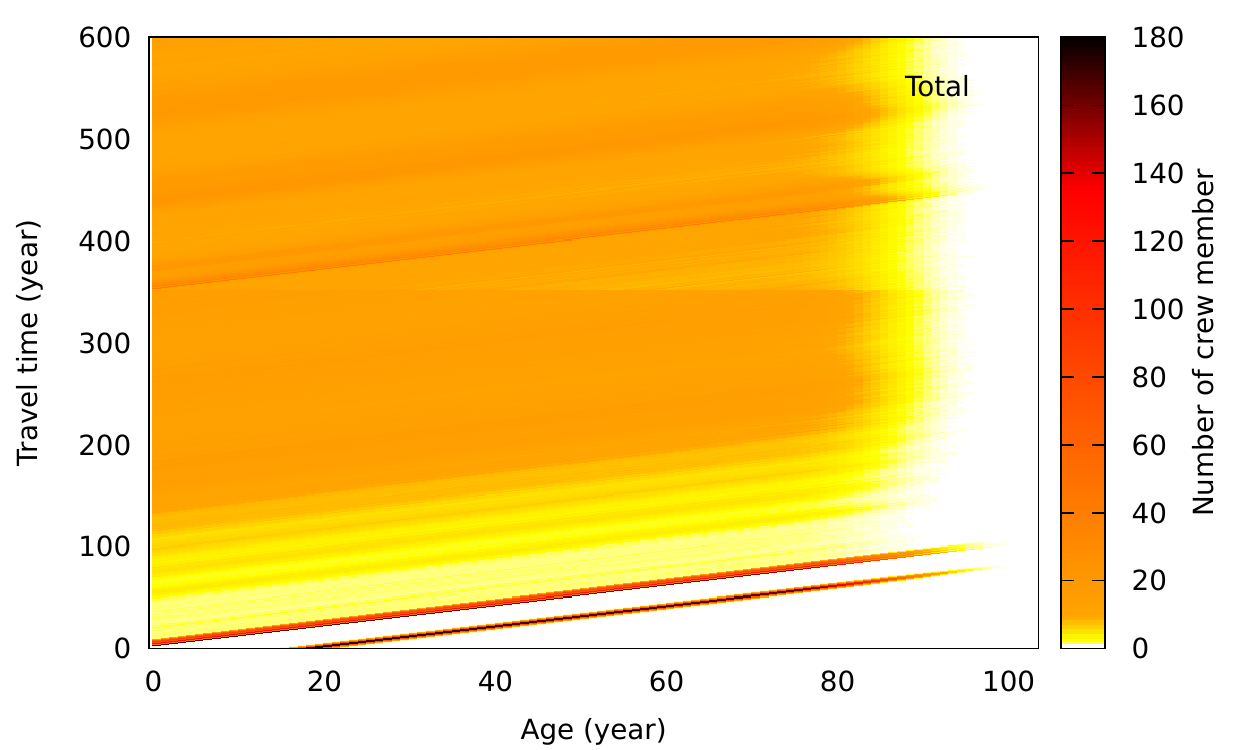}
    \caption{Density of crew members of a given age (x-axis) over time (y-axis). 
	    The population density is color-coded from 0 (white) to maximum (black).}
  \end{subfigure}

  \medskip

  \begin{subfigure}[t]{.48\textwidth}
    \centering
    \includegraphics[width=\linewidth]{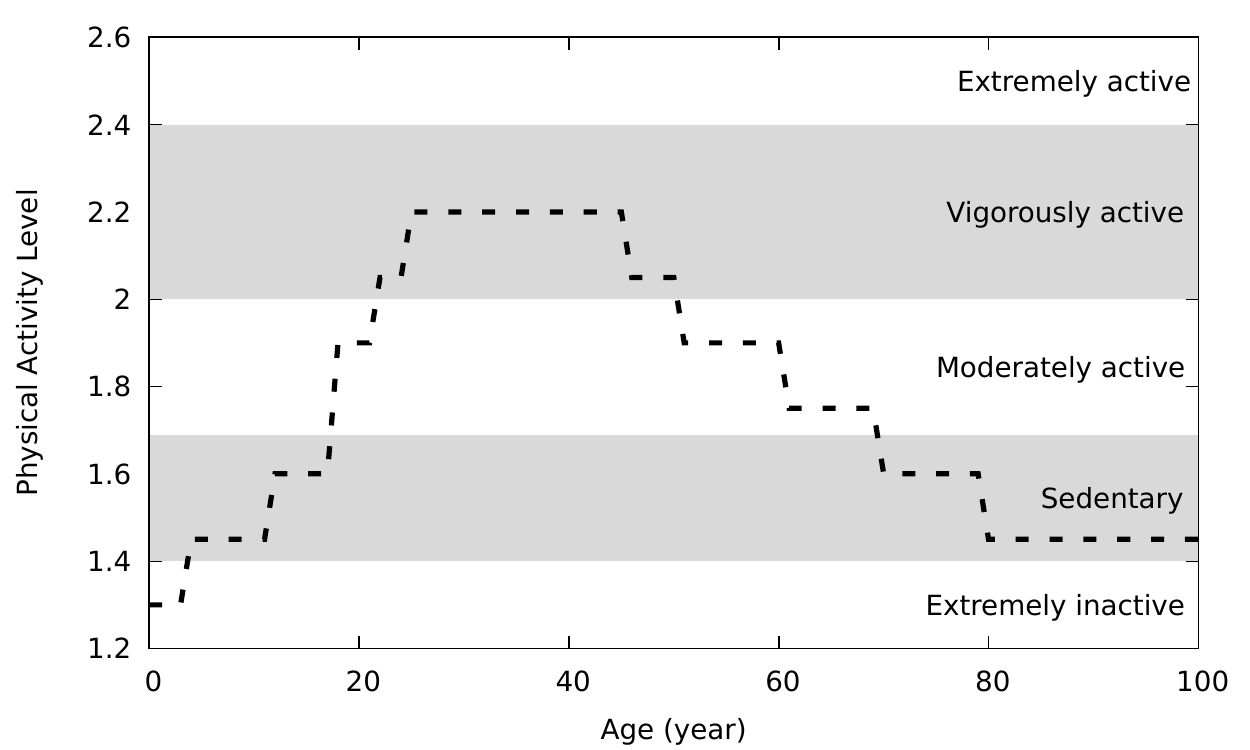}
    \caption{Physical activity level (PAL) scenario for our population model.}
  \end{subfigure}
  \hfill
  \begin{subfigure}[t]{.48\textwidth}
    \centering
    \includegraphics[width=\linewidth]{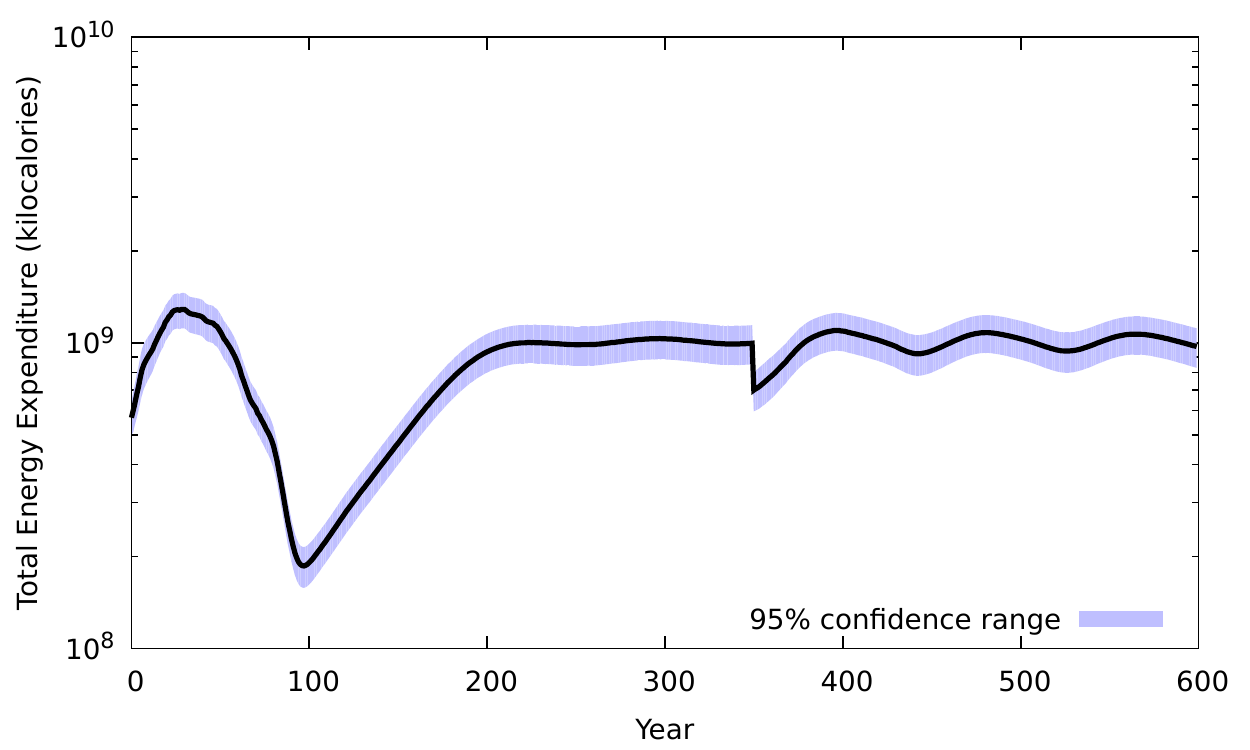}
    \caption{Total energy expenditure (TEE, in kilo-calories) per year in the 
	     vessel for the considered PAL scenario.}
  \end{subfigure}
  
  \caption{HERITAGE results for a 600 years-long interstellar travel under 
	  the conditions described in the text.}
  \label{Fig:Simulation}%
\end{figure*}

In Fig.~\ref{Fig:Simulation}, we present the outcomes of the simulation in terms of demography and food consumption on-board. We can see that 
the population quickly heads towards the overpopulation threshold in two generations. The adaptive social engineering principles activate and 
we see a decrease of the population demography before a slower, second increase that would ultimately lead to stabilization except for the 
presence of the catastrophe at year 350. The strong demographical decrease at year $\sim$ 50 is due to the presence of sibships within the 
spacecraft: the zeroth-generation being very young, its results in well-defined demographic echelons during the first 200 years such as 
seen from Fig.~\ref{Fig:Simulation}~(b). Those demographic echelons, predicted by Moore \cite{Moore2003}, are impacting at the beginning of 
the travel until the various generation clusters mix in age. We also note that the minor demographic changes compared to the previous
papers have not had a significant change on the end results. The physical activity level scenario in our simulation (Fig.~\ref{Fig:Simulation}, c)
drives a total energy expenditure of approximately 10$^9$ kilo-calories per year in the vessel (Fig.~\ref{Fig:Simulation}, d). From a demographic 
and metabolic point-of-view, the crew seems perfectly fine at the end of the mission.

\subsection{Genetic results}
\label{Model:Genetics}

\subsubsection{The Hardy-Weinberg equilibrium}
\label{Model:Genetics:HW}

How to verify the genetic health of the multi-generational crew? The first test to be carried out is whether the population is at the 
Hardy-Weinberg (HW) equilibrium when there are no mutations \cite{Hardy1908,Weinberg1908}: for a sufficiently large population (ideally infinite), 
the frequency of alleles (for non-sexual chromosomes) should tend to be stable over long periods (in our case $>$ 600 years). If the frequencies 
stay almost constant, this would confirm that the initial genetic diversity is likely to remain more or less constant (the population 
remains genetically varied like the original one). This means that the number of breeding individuals is sufficient to ensure a constant
mixing of genes. This verification makes it possible to say that, if there are no mutations (and therefore no increase in potential 
genetic variability), the population is likely to be genetically stable. This is the first test geneticists would use before a 
multi-generational trip to probe the potential stability of the population.

\begin{figure}[!t]
\centering
  \includegraphics[trim = 0mm 0mm 0mm 0mm, clip, width=8.2cm]{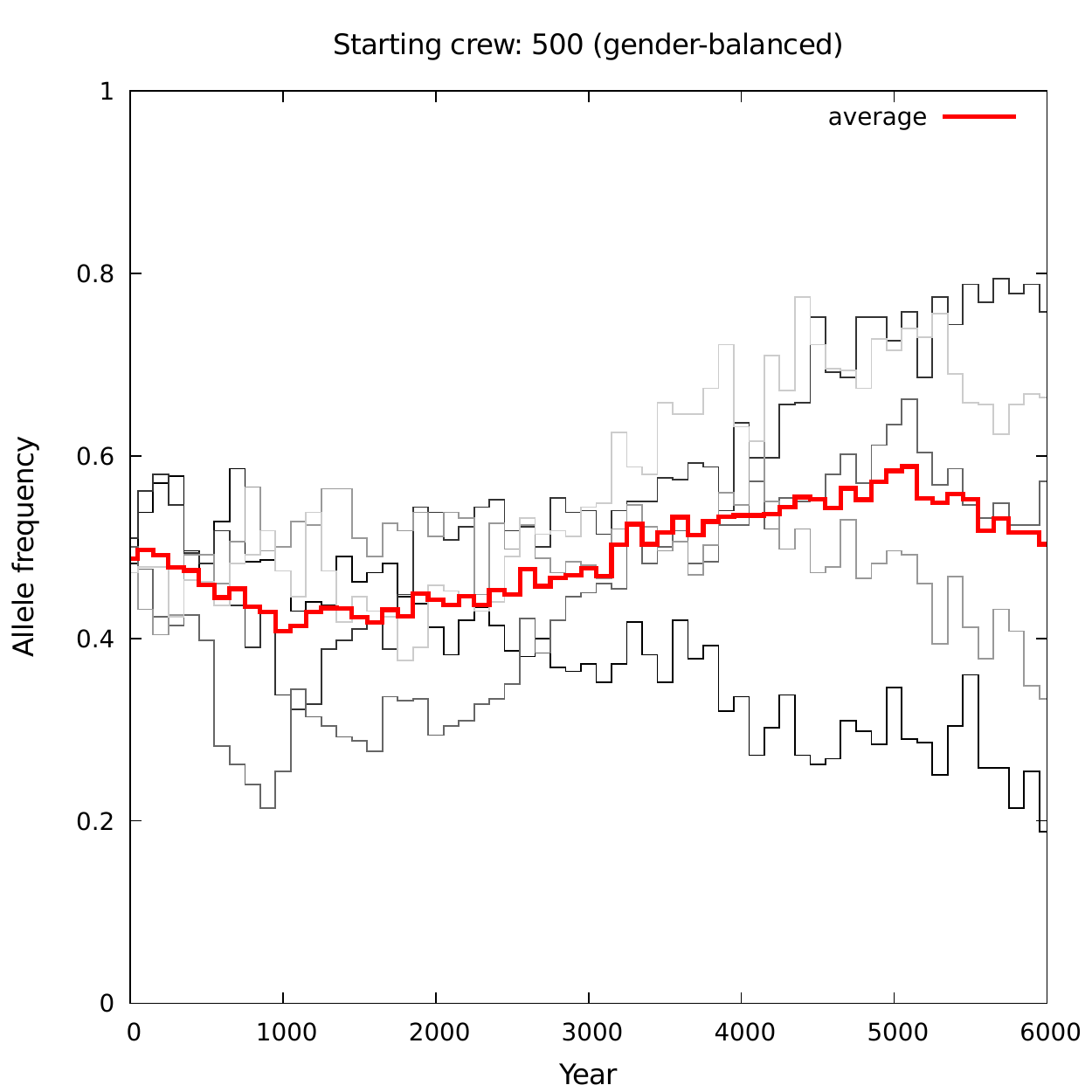}
  \caption{Results of genetic drift on the frequency of an allele at a randomly-chosen locus (that 
	   has two allelic forms) in a randomly-chosen chromosome. The changes in frequency is followed 
	   at every generation from the initial population presented in Tab.~\ref{Tab:Parameters}. 
	   The five gray lines represent the frequency of an allele in 5 independent populations 
	   while the thick red line represents the average frequency. The simulation was extended to 
	   6\,000 years in order to see any long-term effect on the genetic stability of the populations. 
	   The initial population consists of 500 gender-balanced individuals.}
  \label{Fig:HW_population}
\end{figure}  

\begin{figure}[!t]
\centering
  \includegraphics[trim = 0mm 0mm 0mm 0mm, clip, width=8.2cm]{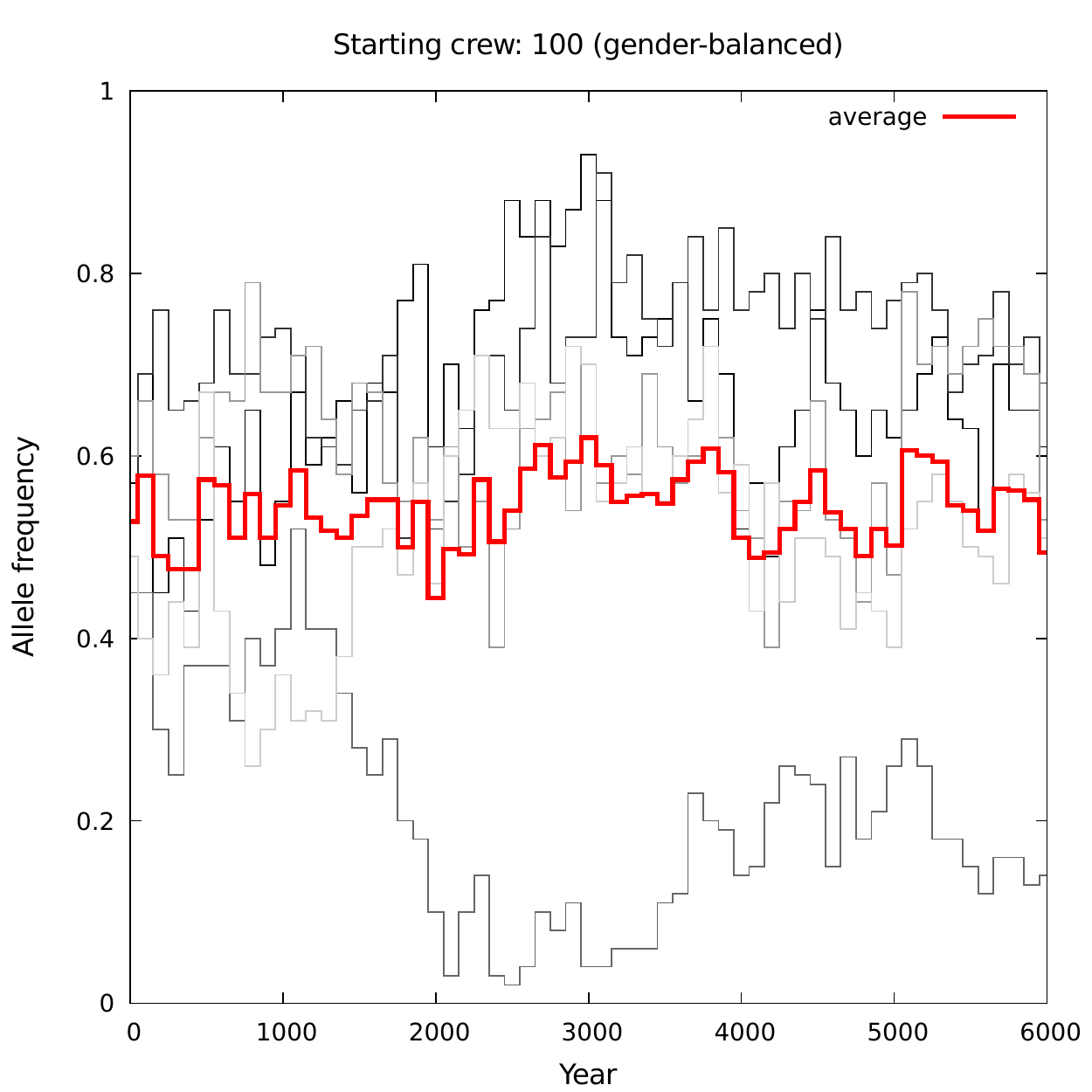}
  \caption{Same as in Fig.~\ref{Fig:HW_population} but for an initial crew
	   of 100 gender-balanced persons. This smaller population allows us 
	   to visualize the effects of the starting population's size on the 
	   genetic drift.}
  \label{Fig:HW_population_100}
\end{figure}  

In Fig.~\ref{Fig:HW_population}, we explored the allele frequency variations inside the multi-generational population with a departing 
crew of 500. To do so, we artificially imposed the frequency of the alleles at a random locus along the genome to be randomly but equally distributed 
between integer values 0 and 1. In other words, a randomly chosen locus is assigned two possible alleles with frequencies of 0.5 (50\%) each. This is 
an idealized and simplified situation that facilitates the visualization of the genetic drift on the frequency of one of both alleles to determine whether 
this specific locus follows the Hardy-Weinberg equilibrium (stable frequency of alleles over time). We then ran HERITAGE and increased the duration of 
the interstellar travel from 600 to 6\,000 years to check whether dramatic allele frequency variations could happen after the nominal period of 600 years. 
For the initial crew described in the previous section, we find that the average of numerous (repeated) simulations tends to show a nearly-constant 
frequency around 50\%, indicative of the fact that the frequency of the allele under inspection remains stable over time, i.e. is nearly at the Hardy-Weinberg 
equilibrium, which is what one would expect for a population of more than 50 people \cite{Keats2013}. The differences are simply due to the fact that we 
do not necessarily have clearly separated generations, and that the individuals are not all synchronous (in births, in age, in reproduction timing, etc.)
contrary to theoretical models \cite{Edwards2008}. From these deviations from HW conditions, small stochastic variations of allelic frequencies occur even 
in the case of a population theoretically composed of enough individuals (more than 50 reproducing individuals). For a smaller population (Fig.~\ref{Fig:HW_population_100}), 
sampling bias and deviations from the theoretical HW conditions more strongly affect allele frequencies. The averaged frequency does also oscillate around 
50\% but stochastic variations are much larger. Yet, the population is close to the Hardy-Weinberg equilibrium, although it is more easily affected by 
sampling bias, which means that the HW equilibrium is more easily lost with a smaller population and that larger populations are better at
avoiding this. This was already expected from previous findings \cite{Keats2013}, from results obtained by monitoring the heterozygosity index (see above) 
and it advantageously confirms the conclusions of our previous publication, 
in which we stated that at least 98 people should constitute the zeroth-generation crew of any multi-generational mission \cite{Marin2018}. Our first preliminary 
tests, applied to one bi-allelic locus, indicate that our multi-generational population would likely be genetically healthy, in the sense that 100 to 500 people would
be enough to stabilize allele frequencies and, consequently, the starting genetic diversity, at least over a 6\,000 year-long period of time. Of course, now 
that we have implemented virtual genomes that carry thousands of loci with multiple possible allelic forms, we will be able to extend this analysis (HW equilibrium, 
frequency variations, etc.) at the genome-scale with more appropriate tools as we did by measuring polymorphism, heterozygosity indices and consanguinity.

\subsubsection{Nei's minimum genetic distance}
\label{Model:Genetics:Nei}

Now that we have confirmed that the allele frequencies in the crew are likely stable in the long run, it is necessary to determine the impact of 
time (i.e. genetic recombination and shuffling during meiosis, matings, etc.) on the genetic composition of the crew in comparison to the zeroth-generation. 
To do so, we can compute the genetic distance $D_{\rm A}$ that measures the degree of differentiation: populations with many similar alleles have small 
genetic distances (they share similar genotypes, i.e. allelic patterns and are closer to a ``common ancestor'' population) while populations with more 
different allelic patterns or genotypes are separated by greater genetic distances. To determine the genetic distance, we used the Nei's minimum genetic 
distance \cite{Nei1974} that assumes that genetic differences arise mainly from mutation and genetic drift (which is the case here):

\begin{equation*}
  \textsl{\textrm{D}}_{\rm A} = 1 - \sum_{l}\sum_{u}\sqrt{X_u Y_u} / L ,
\end{equation*}

where $X$ and $Y$ represent two different populations for which $L$ loci have been studied. In our calculus, $X$ is the zeroth-generation while 
$Y$ is the population after a given time. $X_u$ and $Y_u$ represent the $u^{th}$ allele frequencies at the $l^{th}$ locus. $D_{\rm A}$ = 0 means 
that the n$^{th}$ generation is identical to the zeroth-generation and with increasing $D_{\rm A}$'s the genome of the population starts to differ
from the initial one. A Nei's minimum genetic distance between 0 and 0.05\% indicates that the initial and final populations are very similar (in the sense 
that they share very similar allelic patterns) and likely poorly differentiated \cite{Nei1976}. A value between 0.02 and 0.2\% usually indicates that 
two (or more) populations are likely subspecies (e.g., the Bengal and Siberian tigers are examples of subspecies), which should be 
understood as ``populations with sufficiently different allelic/genetic patterns to arbitrarily subdivise them into distinct entities'' \cite{Nei1976}. 
Values between 0.1 and 2\% usually imply that the two (or more) populations under investigation are different species (e.g., cats, chickens and chimpanzees 
are three examples of species) \cite{Nei1976}. Here, the term ``species'' must be considered carefully. In general, individuals or populations are considered 
to belong to the same species if they can breed and produce descendants that are themselves inter-fertile. However, in an evolutionary point-of-view, the 
concept of species arbitrarily encompasses all individuals that belong to a continuous genealogical flow/continuum (meaning that time is also part of the definition). 
All groups of individuals (populations) that have, though, similar genetic features but that do not (or not anymore) contribute to this genealogical flow 
(infertility, incompatibility of gametes, etc.), and that are part of a sister, but separated genealogical flow, are considered other species, 
by definition. Speciation is therefore not a discrete event, since the assignment of individuals to a particular group is arbitrary. However, 
the emergence of reproductive and/or biological incompatibilities (genetic reproductive barriers, etc.) constitutes the origin of separated genetic lineages
that we name species \cite{Galtier2018}. The amount of genetic differences required to reproductively isolate populations from each other is not known and depends on incalculable
possible combinations of genetic (mutational, genotypic, etc.), phenotypic (including behavioral) and environmental effects. For this reason, and because we
did not (and cannot) model the tremendously complex mechanisms that drive reproductive compatibility, or influence reproductive isolation in the case of 
isolated human populations traveling through space for generations, the evaluation of the genetic distance (following Nei) will not be used to speculate on
the emergence of human subspecies or species.

\begin{figure}[!t]
\centering
  \includegraphics[trim = 0mm 0mm 0mm 0mm, clip, width=8.2cm]{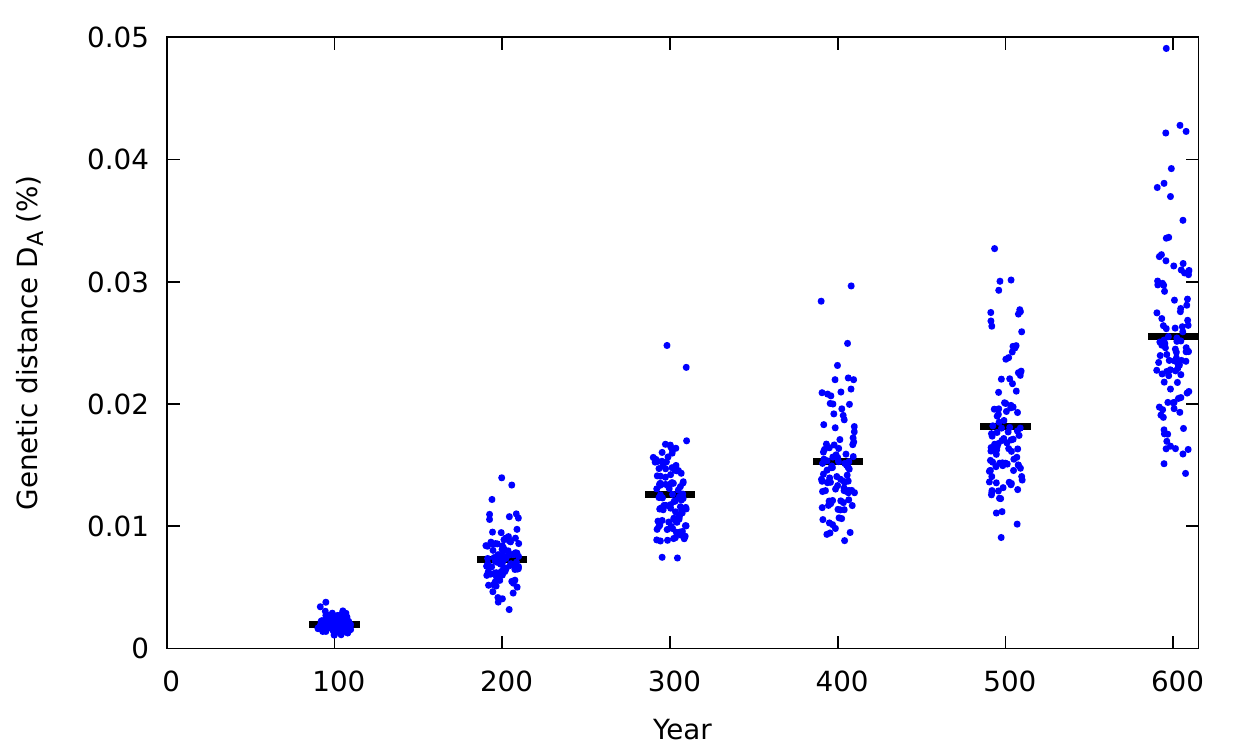}
  \caption{Evolution of the Nei's minimum genetic distance $D_{\rm A}$
	   as a function of time for the HERITAGE parametrization presented 
	   in Fig.~\ref{Fig:Simulation}. Each point (at a given time-step) 
	   represents the outcome of one out of one hundred simulations. 
	   The averaged genetic distance is highlighted by a black bar.
	   The annual equivalent dose of 
	   cosmic ray radiation is similar to the Earth radioactivity 
	   background at sea level (0.3~mSv).}
  \label{Fig:Genetics}
\end{figure}    

\begin{figure}[!t]
\centering
  \includegraphics[trim = 0mm 0mm 0mm 0mm, clip, width=8.2cm]{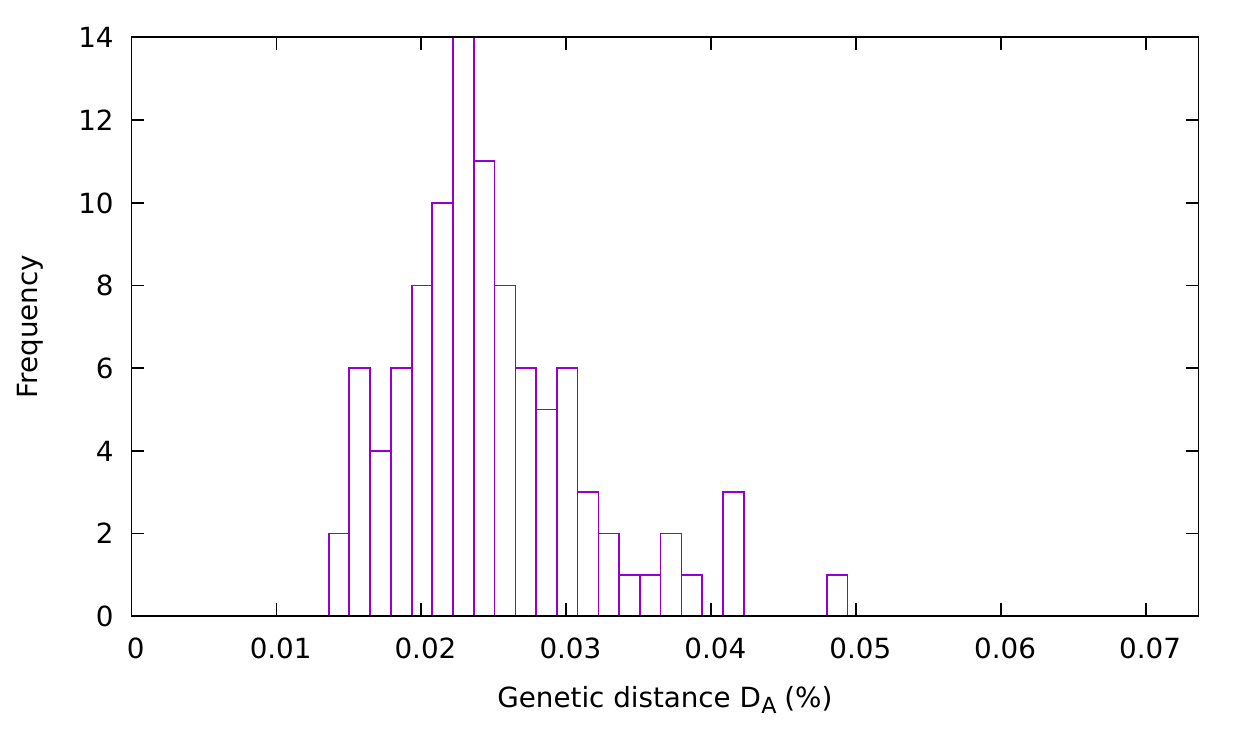}
  \caption{Histogram of the Nei's minimum genetic distance 
	    between the final 100 populations and their 
	    respective zeroth-generation genome after 600 
	    years of interstellar travel with low cosmic 
	    ray radiation (0.3~mSv).}
  \label{Fig:Histo_Genetics}
\end{figure}

We plot in Fig.~\ref{Fig:Genetics} the time-dependent evolution of the genetic distance between the zeroth and the n$^{th}$ generations (with 100 
years steps). Each point (at a given time-step) represents the outcome of one out of one hundred simulations. The averaged genetic distance
$D_{\rm A}$ is highlighted by a black bar. We can see a large spread around the mean that is predominantly due to the effect of genetic drift, 
that can change allelic patterns (haplotypes, genotypes). The increase in genetic distance is somewhat linear with time if we consider the mean 
values (such as expected from \cite{Nei1976}) but we also observe that the tail of the $D_{\rm A}$ distribution is quite large. This means that 
the outcomes of 600 years of space travel most often lead to genetically not-so-different populations with respect to their zeroth-generation. 
This is in agreement with the results presented above, from which we concluded that our traveling populations are close to the HW equilibrium, 
a state that ensures the stabilization of alleles within populations. Consequently, under the neutral hypothesis condition we used (no phenotypic
effect of mutations, allelic patterns, etc.), no natural selection of alleles or allelic patterns (haplotypes or genotypes) are expected, which 
reduces the otherwise unavoidable genetic differentiation of the population, with respect to the initial one (or to the one that remained on Earth). 
Yet, the genetic distance increases with time, because the deviations from the Hardy-Weinberg conditions, although small, tend to have cumulative 
effects that eventually affect the genetic composition of populations. Genetic differentiation, as expected, is unavoidable, because the genetic 
input within the vessel is limited, subjected to genetic drift with potential losses of alleles, and there is no external genetic input (from 
other human populations) that could replenish the genetic pool with primitive (original) alleles that could reconstitute the starting allelic 
diversity. Some extreme cases ($D_{\rm A} \approx$ 0.05\%) may appear from those purely stochastic sampling effects (recombination, chromosome 
shuffling, mating), leading to genetically different sub-groups on-board (with respect to the 0$^{th}$ generation). The distribution of the Nei's 
minimum genetic distance at the 600 year time-step is illustrated in Fig.~\ref{Fig:Histo_Genetics}. It is a positively skewed unimodal distribution 
(the tail is on the right of the histogram) that originates from the relative small size of the populations but also from genetic drift. In some 
rare and extreme cases, several allele frequencies can drastically change, affecting the allelic composition, leading to larger genetic distances. 
Note that genetic differentiation occurs on-board during the journey, and that is will continue after arrival. As mentioned, measuring the Nei's 
distance does not tell anything on the speciation of interstellar populations, but certainly illustrates that genetic differentiation will occur 
at rates that depend on the genetic drift and on the populations' size.

\begin{figure}[!t]
\centering
  \includegraphics[trim = 0mm 0mm 0mm 0mm, clip, width=8.2cm]{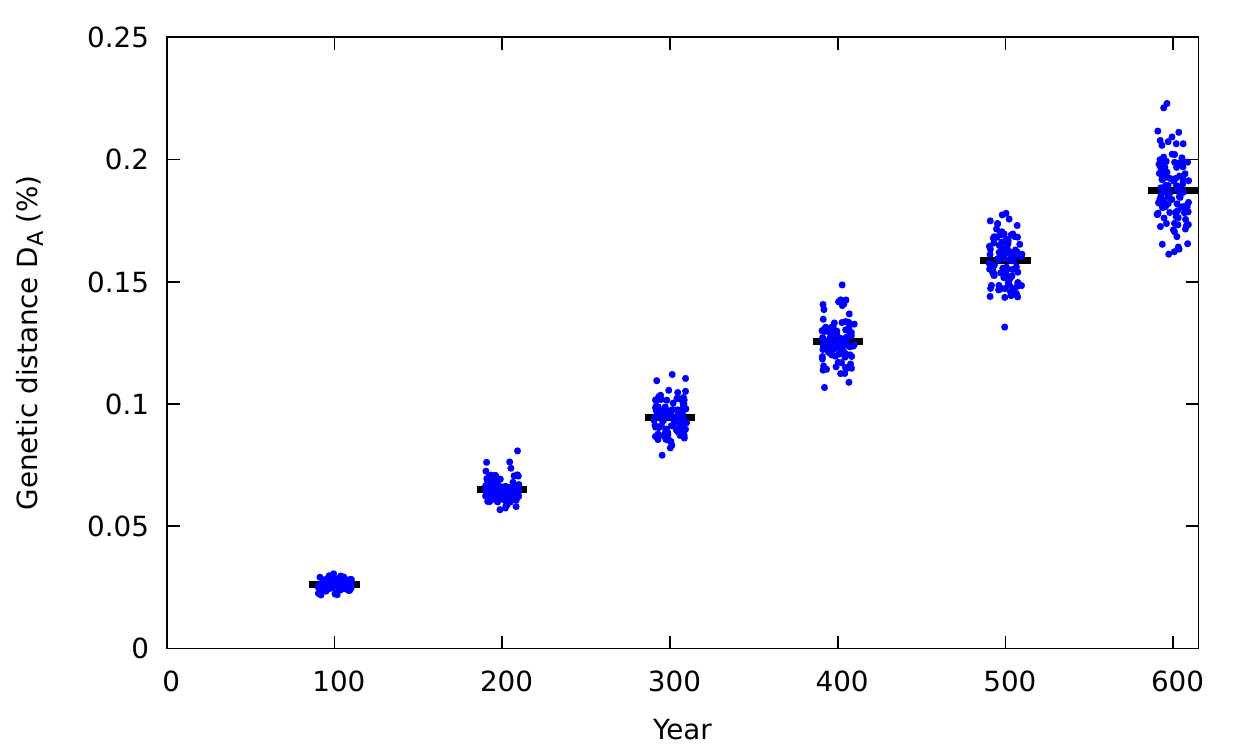}
  \caption{Same as Fig.~\ref{Fig:Genetics} but with a higher 
	  annual equivalent dose of cosmic ray radiation (300~mSv).}
  \label{Fig:Genetics_300mSv}
\end{figure}

\begin{figure}[!t]
\centering
  \includegraphics[trim = 0mm 0mm 0mm 0mm, clip, width=8.2cm]{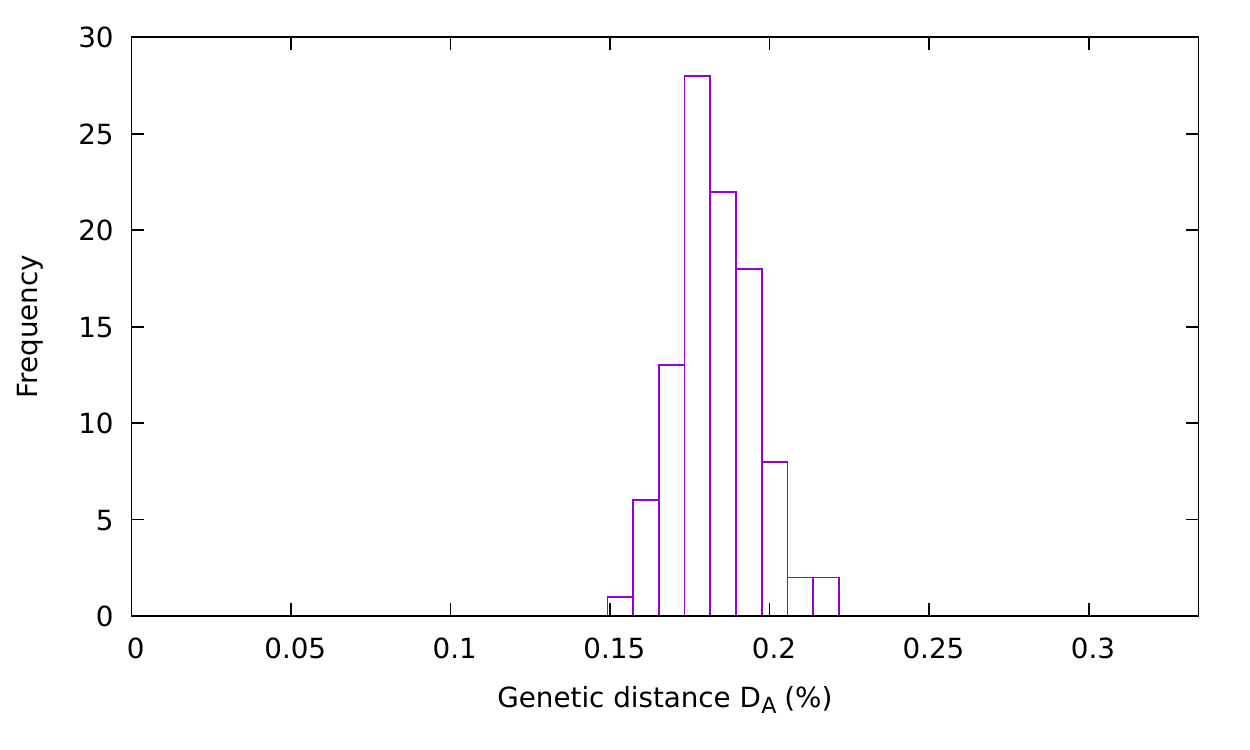}
  \caption{Same as Fig.~\ref{Fig:Histo_Genetics} but with a higher 
	  annual equivalent dose of cosmic ray radiation (300~mSv).}
  \label{Fig:Histo_Genetics_300mSv}
\end{figure}

For a purely scholastic experience, we can run one hundred more simulations with a much higher annual equivalent dose of cosmic ray radiation (300~mSv).
While the crew would likely be wiped out by cancers and genetic disorders first, we can observe in Figs.~\ref{Fig:Genetics_300mSv} and \ref{Fig:Histo_Genetics_300mSv}
the effects of neutral mutations onto the Nei's genetic distance after 600 years under those extreme conditions. In Fig.~\ref{Fig:Genetics_300mSv} we 
see that speciation (in the sense of Nei) or, more properly formulated, ``strong genetic differentiation'' would occur relatively quickly ($\ge$ 300 years) 
but also that the spread in genetic distances between the one hundred different populations is not so large. This is due to the fact that genetic drift 
becomes less important than the accumulation (at high rates) of spontaneous mutations. This latter phenomenon is unlikely to occur in reality, due to 
selection effects that would filter out numerous mutations or allelic combinations, reducing the differentiation rate accordingly. The histogram of 
$D_{\rm A}$ at year 600 (Fig.~\ref{Fig:Histo_Genetics_300mSv}) is still a positively skewed unimodal distribution but the skewness factor is less 
prominent than in the case of negligible radiation doses. 

\section{Conclusions and further development}
\label{Conclusions}

We have significantly upgraded the agent based Monte Carlo code HERITAGE in order to include a representative toy model of the human 
genome for each crew member. We implemented biologically realistic gamete production processes (meiosis), including crossing-over, 
unilateral conversion, chromosomes and chromatids shuffling, etc. Those new implementations allow us to perform genetic simulations 
on multi-generational populations in a large parameter phase space. We can now determine if, from an initial population of a given size 
and with a defined genetic composition, it would be possible to preserve sufficient allelic diversity over time and test the contribution 
of neomutations from cosmic ray radiation. 

In this work, we assumed that all combinations of alleles (genotypes) and mutations have neutral phenotypic effects (neutral hypothesis). 
In this case, and using a single bi-allelic locus as a probe (as in \cite{Smith2014}), we found that a MVP of about 100 gender-balanced 
people at the beginning of the interstellar travel would likely be close to the Hardy-Weinberg equilibrium (during the course of a 6000 years 
journey), which is expected to preserve most of the genetic diversity selected for the initial crew. This initial crew can also sustain 
genetic drift and small amounts of (neutral) neomutations, and arrive ``genetically healthy'' at the end of 600 years of deep space travel 
with state-of-the-art radiation shields. However, we observed that the Hardy-Weinberg equilibrium can easily be lost, and that its 
stabilization effect suppressed in the case of stronger radiation fluxes or catastrophic demographic events. It results that a safer and 
more adapted population threshold should be considered, probably in the range of a few hundred individuals. We indeed have experimented our
code with a MVP of 500 crew members and found it more resilient. Nevertheless, it is too soon to conclude yet since we must exploit
the full potential of our code to analyze genetic effects (HW equilibrium as a function of the number of crew members and allelic diversity) 
at the genomic (multi-locus) scale. We also have to move away from the neutral hypothesis and include phenotypic effects of allelic combinations, 
and more importantly, of neo-mutations, since this is likely to strongly influence the evolution of the genetic composition and structure of 
populations. Indeed, as discussed above, allelic combinations, mutations and interactions with the environment (external or internal, i.e. 
biological, cellular, etc.) can lead to selective effects (positive, negative or neutral effects). This will be achieved in the second part 
of this paper series. To take into account these complex effects, we will rely on the extended literature on population genetics of mutations 
\cite{Kimura1978,Fudala2009,Masel2013}. It is highly probable that strong annual doses of radiation will most likely wipe out the interstellar 
crew rather than lead to strong differentiation in a time-frame of a few hundred years. However, we will be able to check the impact of 
any nuclear accident inside the spacecraft by artificially increasing the annual dose of radiation to a maximum peak for only a year, before 
returning to safer doses. In turn, we will have access to realistic simulations for managing nuclear accidents on Earth, such as the Chernobyl 
catastrophe \cite{Ivanov1997,Davis2004,Zablotska2016} or the more recent triple nuclear meltdown in Fukushima Dai-ichi \cite{Rhodes2014,Taira2014}.

\section*{Acknowledgment}
The authors would like to acknowledge Dr. Rhys Taylor and Ms Esther Collas for 
their comments and suggestions that greatly helped to improve this paper.

\bibliography{mybibfile}

\newpage

\appendix

\section*{APPENDIX A: Validating the numerical biological laws in the code}
\label{Appendix:Test}

\begin{figure*}[!t]
  \begin{subfigure}[t]{.7\textwidth}
    \centering
    \includegraphics[trim = 0mm 12mm 0mm 12mm, clip, width=12cm]{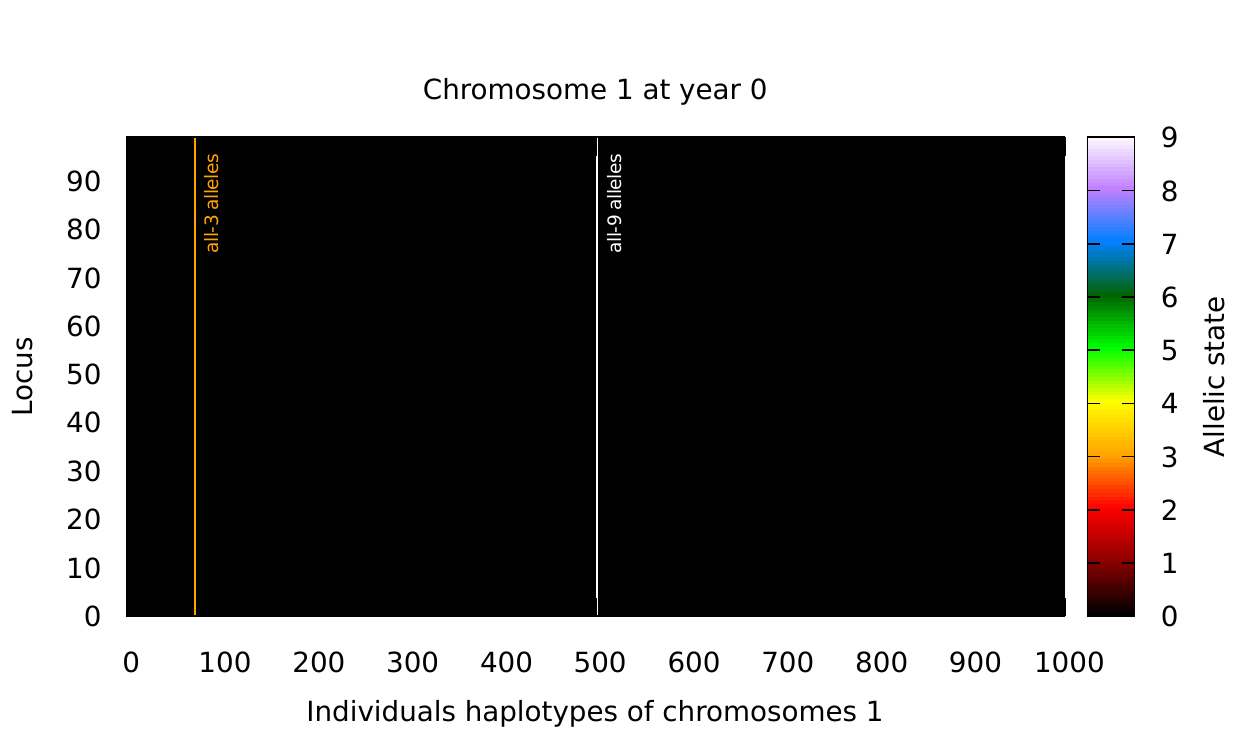}
  \end{subfigure}
  \hfill
  \begin{subfigure}[t]{.3\textwidth}
    \centering
    \includegraphics[trim = 12mm 0mm 30mm 0mm, clip, height=2.2cm, angle=90]{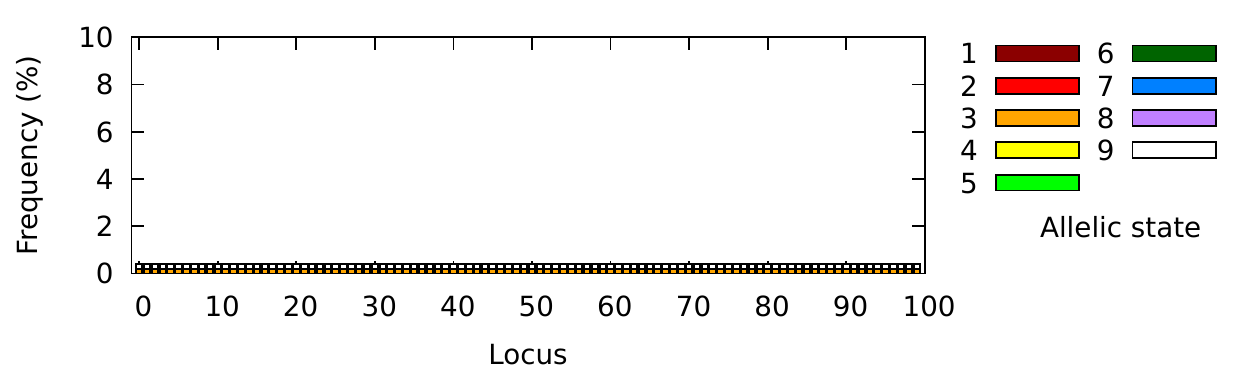}
  \end{subfigure}

  \medskip

  \begin{subfigure}[t]{.7\textwidth}
    \centering
    \includegraphics[trim = 0mm 0mm 0mm 12mm, clip, width=12cm]{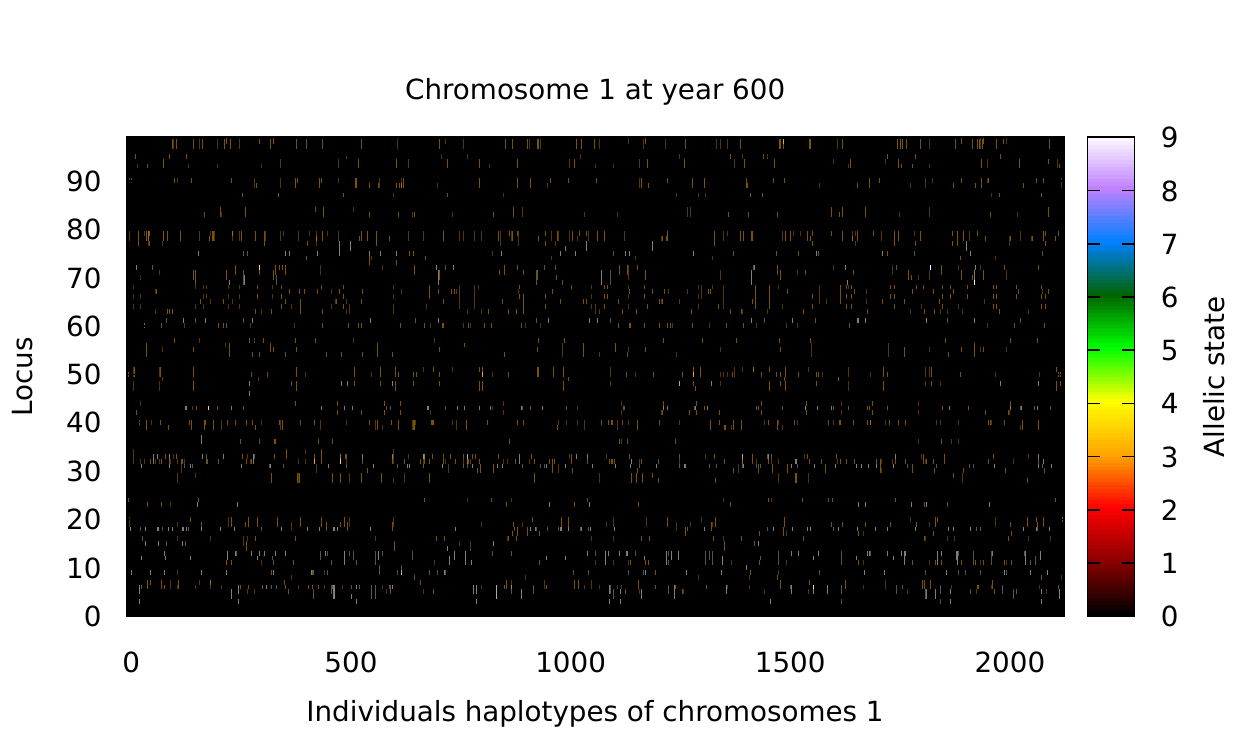}
  \end{subfigure}
  \hfill
  \begin{subfigure}[t]{.3\textwidth}
    \centering
    \vspace{-6.1cm}\includegraphics[trim = 0mm 0mm 30mm 0mm, clip, height=2.2cm, angle=90]{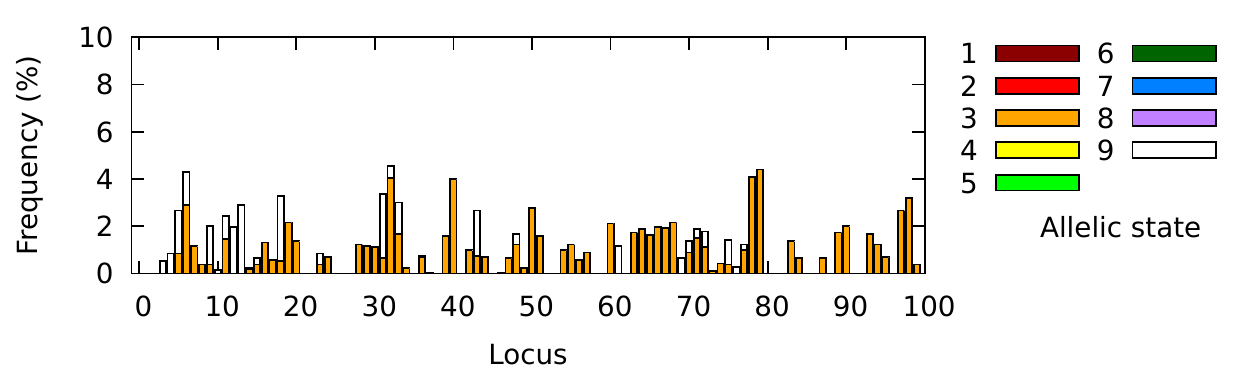}
  \end{subfigure}
  \caption{Haplotypes heat maps of all chromosomes 1 in an initial and final population after 600 years of space travel
  under little-to-no cosmic ray radiation (no mutational effects). The top panel shows the genetic composition of an initial
  theoretical population that is homozygous at all positions. All allelic states are set to 0, except for two individuals
  (``all-3 genotype'' and ``all-9 genotype''). The bottom panel shows the allelic patterns that formed after 600 years of mating
  as a result of genetic recombination, chromosome shuffling, and contingent formation of a novel diploid individual by pooling
  two independent haploid genomes.}
  \label{Fig:Gene_dilution}%
\end{figure*}

We simulated an initial population of 500 individuals (250 males, 250 females) in which all individuals possess a diploid genome with all chromosomes composed of 
loci that all have an allelic state set to 0 (color code: black). Therefore, they are all homozygous at all positions. In this population, one completely homozygous individual has all
alleles set to state 3 (color code: orange, i.e. ``all-3 genotype'') and one second completely homozygous individual has a genotype with all alleles set to state 9 (color code: white, i.e.
``all-9 genotype''). This situation is purely theoretical, but serves as a test case. This population is used as an initial crew in the starship, and simulated for a 600 year-long journey.
The heat map (top panel) presented in Fig.~\ref{Fig:Gene_dilution} shows the 1000 aligned haplotypes of all chromosomes 1 (500 diploid individuals have 2 $\times$ 500 chromosomes 1) in the initial population, 
all shown in black (all-0 state) with only 2 entirely colored in orange (all-3 chromosomes 1) and 2 entirely colored in white (all-9 chromosomes 1) from the two test individuals.
During the interstellar travel, the all-3 and all-9 individuals produced gametes and reproduced with other individuals, transmitting their all-3 and all-9 unchanged haplotypes to their
offspring (recombination and chromosome shuffling occurred during meiosis but since all alleles were identical in the diploid genome, they produced all-3 and all-9 haploid gametes 
respectively). The offspring became entirely heterozygous (one all-3 hapoid genome and one all-0 haploid genome, or one all-9 and one all-0 genome) in every case. These descendants also produced
gametes, that, in this case, produced novel and randomly shuffled combinations of alleles upon meiois, because genetic segments were exchanged between haploid genomes in sexual cells.
At each generation, this random shuffling occurred, with, in addition, the contribution of contingent matings between random male and female individuals to produce diploid descendants
with novel diploid genomes. After 600 years, the haplotype heat map presented in Fig.~\ref{Fig:Gene_dilution} (bottom) shows that discrete genetic segments (containing all-3 or all-9 alleles) originating from 
recombination of the initial all-3 and all-9 haplotypes are still present -- with, however, frequencies that do not exceed 5\% --, but distributed (sliced) across the ~1100 individuals 
living in the vessel. This indicates that the recombination process that we implemented worked as expected, and illustrates how allelic patterns found along chromosomes in starting crew
members can change, recombine, and produce novel patterns as a result of genetic recombination, chromosome shuffling, and contingent formation of a novel diploid individual by pooling
two independent haploid genomes. This shows how biological processes such as meiosis and sexual reproduction introduce stochastic, contingent and random effects that can modify allelic
combinations (haplotypes and genotypes) but also the frequency of alleles in a population. Note that the proportion/frequency of some alleles can increase (as some in our example, see 
Fig.~\ref{Fig:Gene_dilution}), while others can be lost because they are (contingently) not transmitted (same figure).

\section*{APPENDIX B: Measuring the genetic diversity for various populations}
\label{Appendix:Diveristy}

\begin{figure*}[!t]
  \begin{subfigure}[t]{\textwidth}
    \centering
    \includegraphics[width=0.49\linewidth]{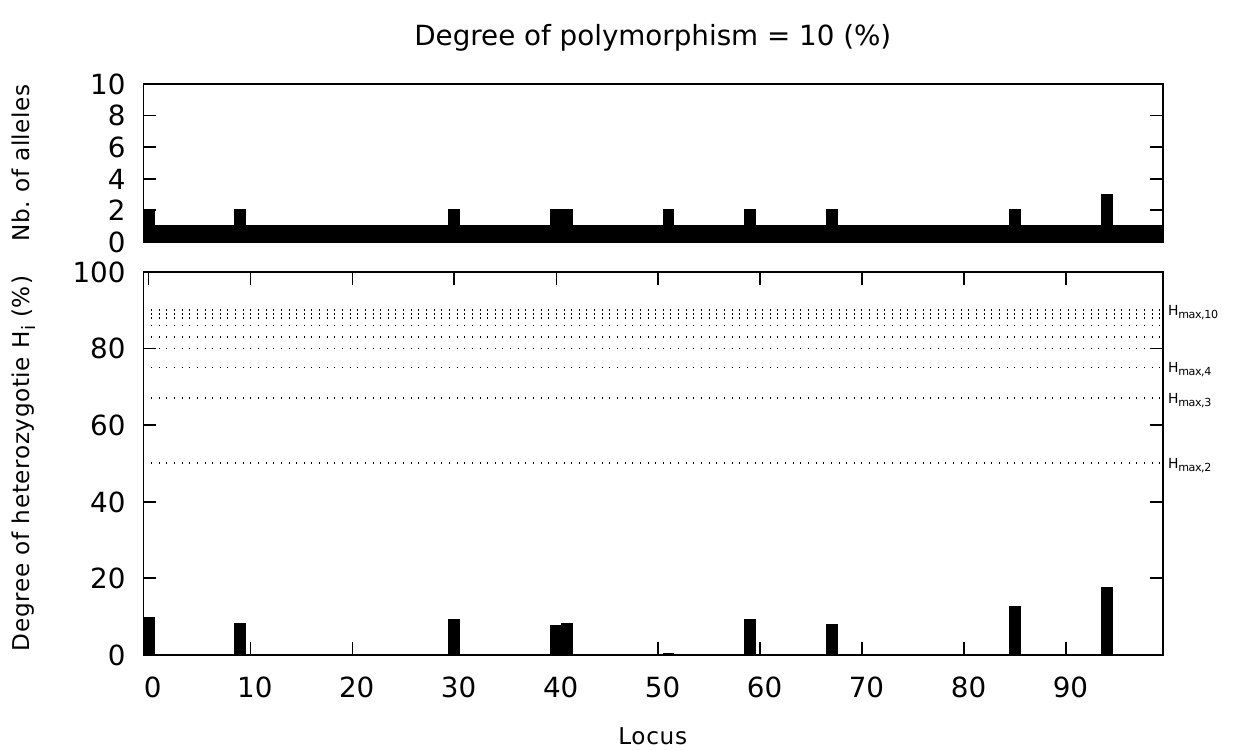}
    \includegraphics[width=0.49\linewidth]{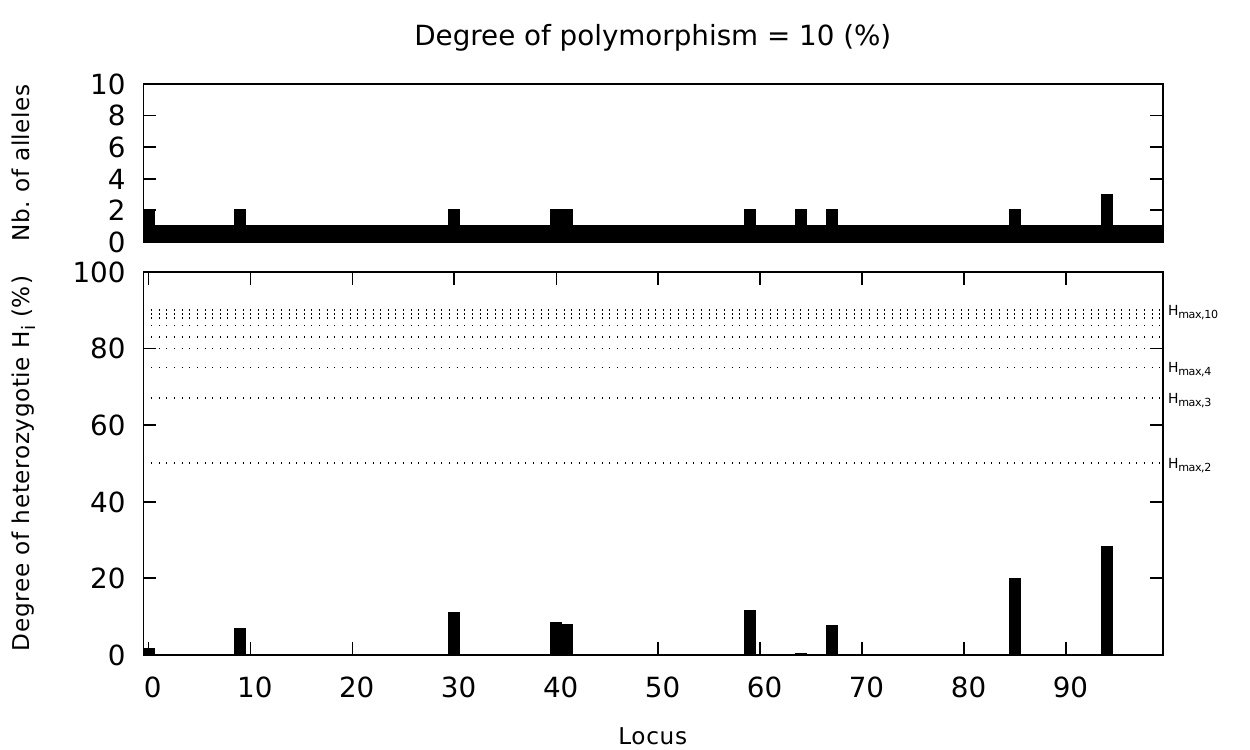}
    \caption{Initial population variation: 0.5\%.}
  \end{subfigure}

  \medskip

  \begin{subfigure}[t]{\textwidth}
    \centering
    \includegraphics[width=0.49\linewidth]{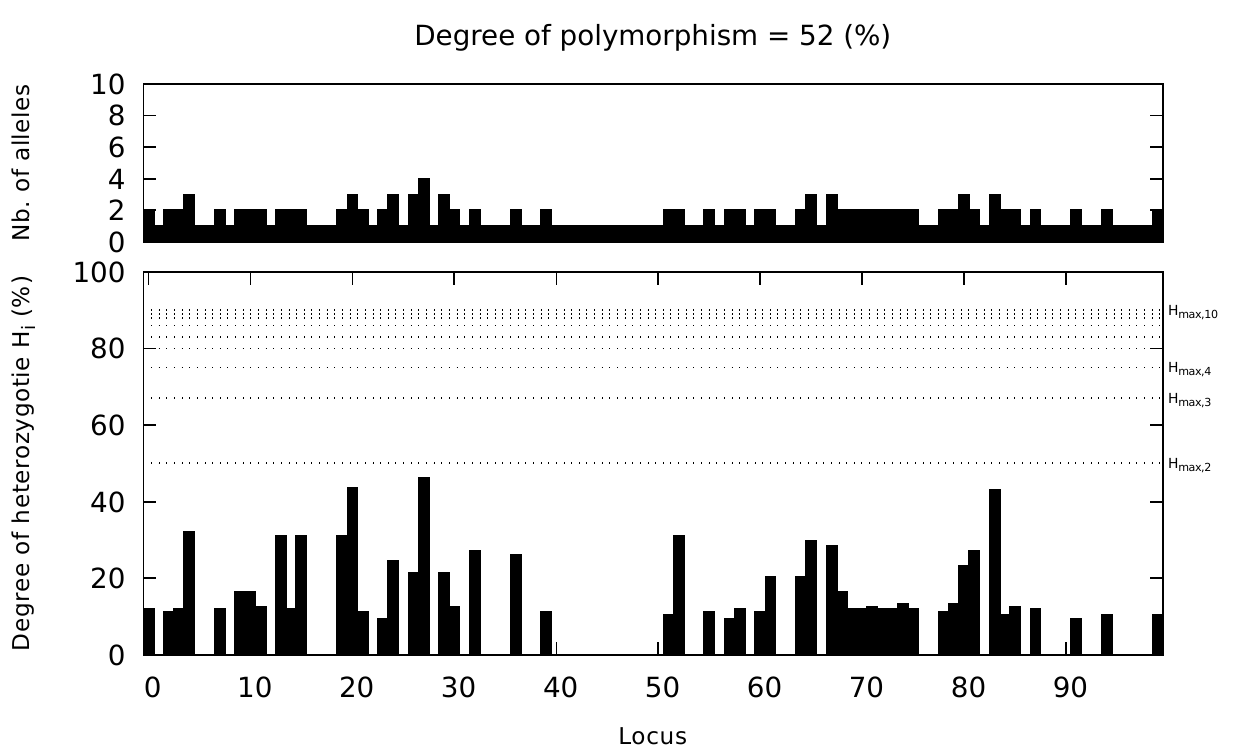}
    \includegraphics[width=0.49\linewidth]{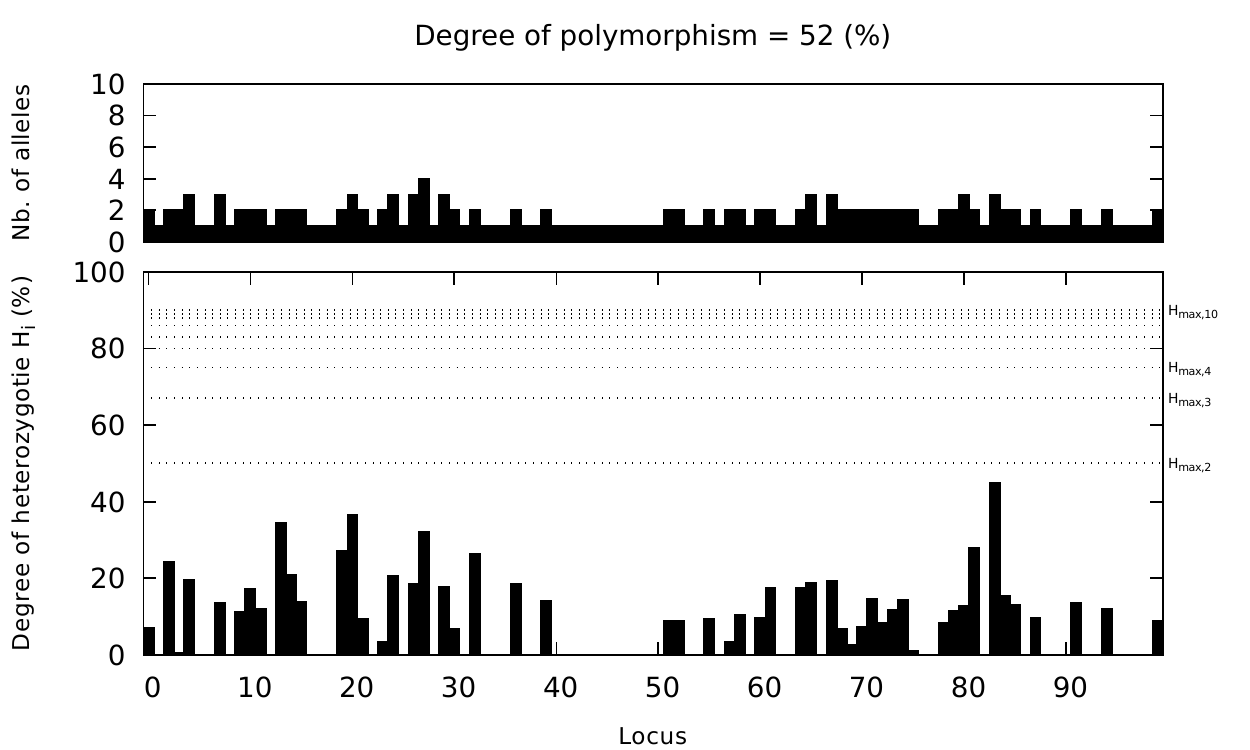}
    \caption{Initial population variation: 5\%.}
  \end{subfigure}

  \medskip

  \begin{subfigure}[t]{\textwidth}
    \centering
    \includegraphics[width=0.49\linewidth]{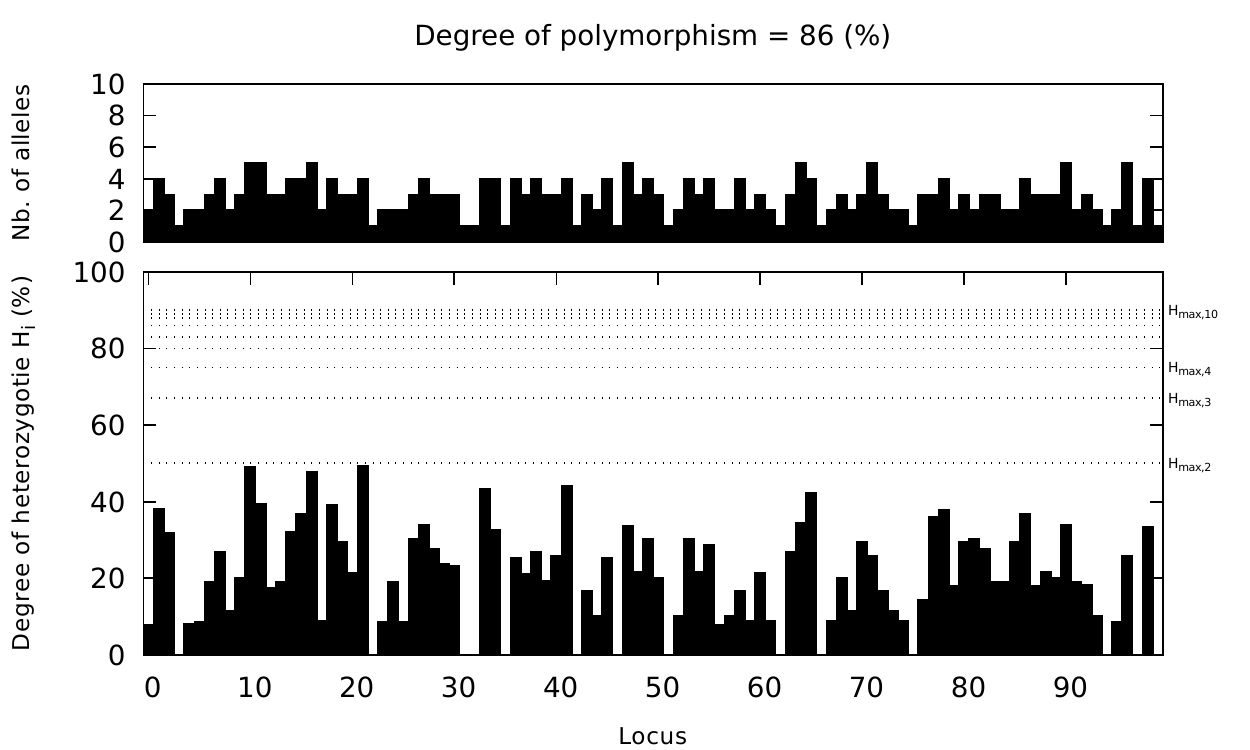}
    \includegraphics[width=0.49\linewidth]{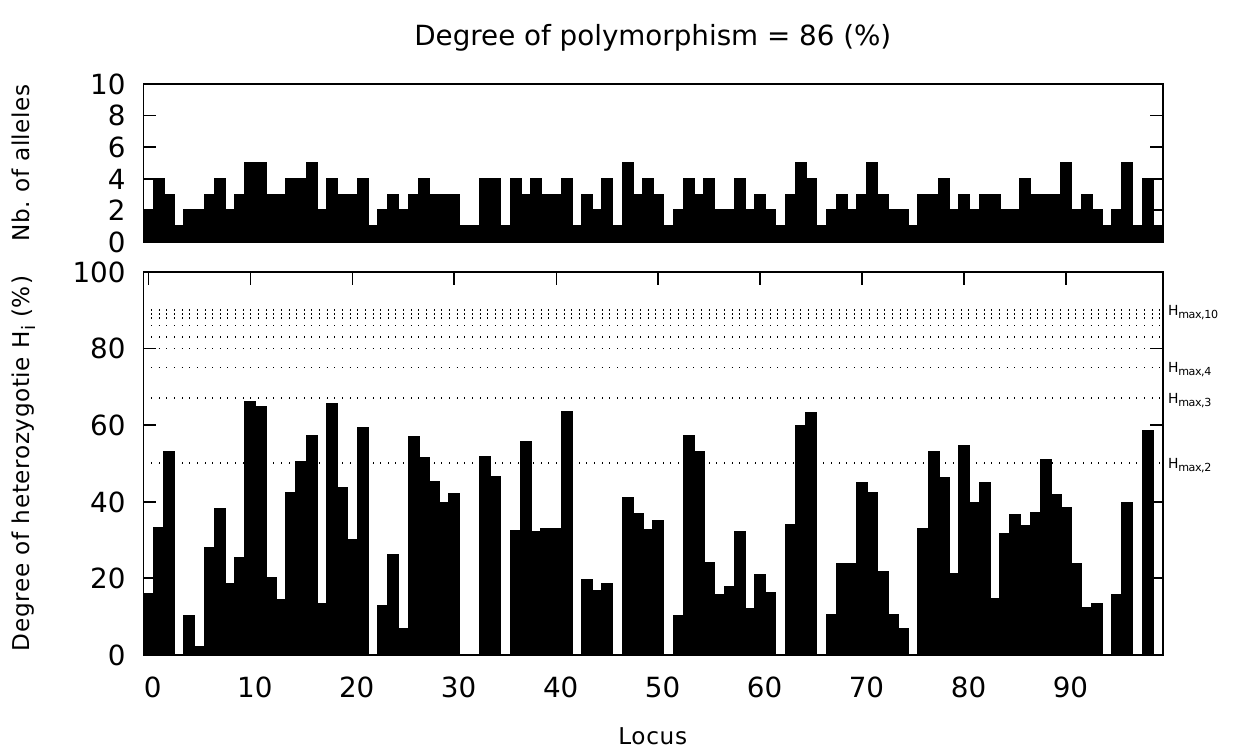}
    \caption{Initial population variation: 20\%.}
  \end{subfigure}  
  
  \caption{Measuring the degree of polymorphism and the heterozygosity index 
	  along chromosome 1. Each figure shows a population whose genomes 
	  show increasing (from top to bottom panels) allelic variations (with
	  respect to the standard reference human genotype). Left column is 
	  at year 0, right column is after 600 years of space travel.}
  \label{Fig:Diversity}%
\end{figure*}

In order to analyze population genetics of the crew during the interstellar journey, we added the possibility to measure several 
parameters that are representative of the genetic diversity within the whole population. The degree of polymorphism (P) indicates 
the proportion of genes within the population (relative to the total number of genes within the reference genome, N), that present
polymorphism, i.e. that can take more than only one allele. If $P$=20\%, this means that 20\% of the 1055 loci ($N$) of the genome
have more than one allelic form. Because P only indicates that a given proportion of genes is polymorphic but does not inform about 
the number of alleles per gene or the relative frequency of those alleles, we also provide a measure of the heterozygosity index 
H$_{i}$ for all loci. H$_{i}$ measures, at position $i$, the proportion of individuals found within the entire population, who are 
heterozygous at this position (have two different alleles at this position in the two haploid genomes at the diploid state). On 
Fig.~\ref{Fig:Diversity}, the number of alleles for each locus is indicated above each corresponding H$_{i}$ value. The H$_{i, max, m}$ thresholds 
are indicated with dot-dashed lines. We remind that H$_{i, max, m}$   indicates the maximal value that H$_{i}$ can take when $m$ alleles
are present at the equilibrium state (equifrequency of alleles). Note that, on Fig.~\ref{Fig:Diversity}, P and H$_{i }$ concern only chromosome~1, 
but HERITAGE calculates P for each chromosome as well as for the entire genome, and H$_{i}$ for all loci.

\begin{figure}[t]
\centering
  \includegraphics[trim = 0mm 0mm 0mm 0mm, clip, width=8.2cm]{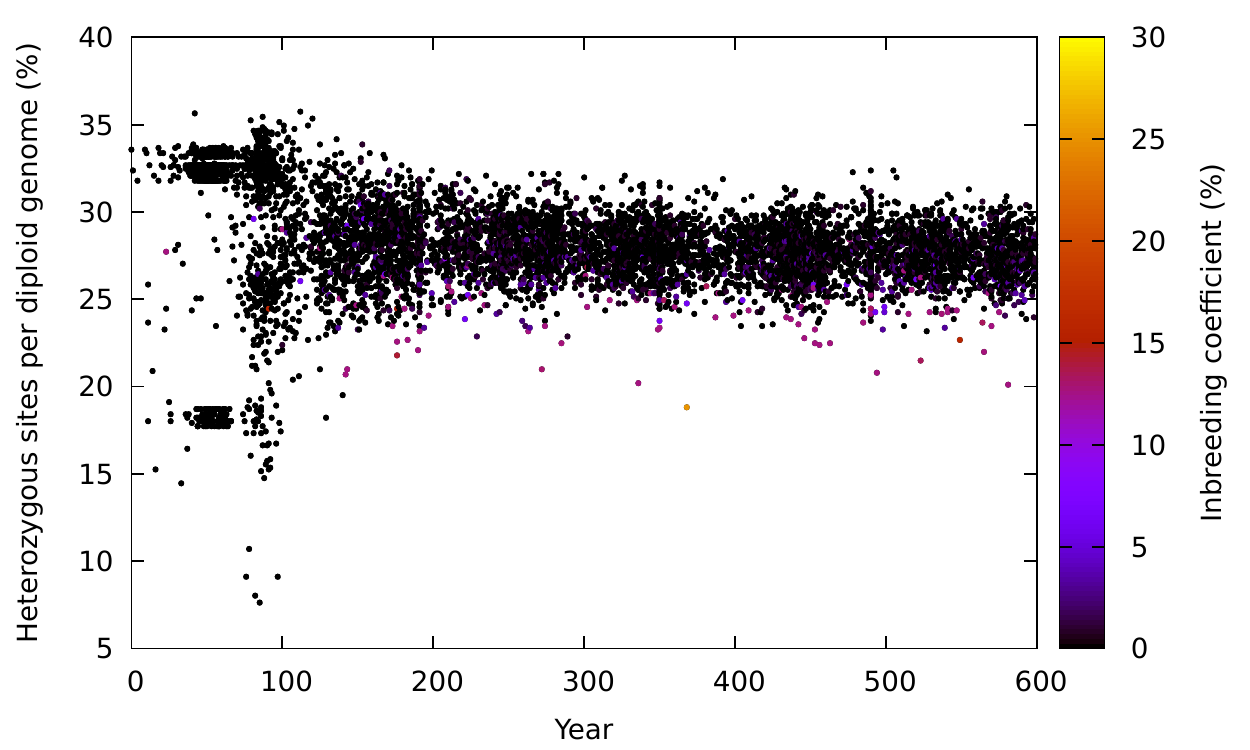}
  \caption{Individual heterozygosity index I$_{k}$ measuring the proportion 
	  of loci that are at the heterozygous state in each individual
	  of the population at the moment of their death.
	  In this simulation, inbreeding was tolerated and the resulting 
	  inbreeding coefficient is shown using a color-code.}
  \label{Fig:Individual_heterozygosity}
\end{figure}

Another way to measure the genetic diversity of the population is to measure I$_{k}$, that we named the individual 
heterozygosity index. I$_{k}$ measures the proportion of loci that are at the heterozygous state in a chosen individual 
(referred to as the k-th individual). In Fig.~\ref{Fig:Individual_heterozygosity}, I$_{k}$ is measured for each individual at the 
moment of death (to account for all possible neomutations) and is calculated for all loci along the entire genome. Each point 
represents an individual and its inbreeding coefficient is indicated using a color code, which enables to appreciate that the lowest 
I$_{k}$ correspond to the highest consanguineous individuals. For a consanguinity factor of 50\% (bother/sister mating), I$_{k}$ 
is expected to drop by approximatively 50\% accordingly. Therefore, in the case of an average I$_{k}$ value of 30\% within the 
population, the most consanguineous individuals are expected to have an I$_{k}$ value of approximatively 15\%. In the case of a 
simulation were consanguineous mating was allowed (such as in Fig.~\ref{Fig:Individual_heterozygosity}), individuals with a 
consanguinity factor of approximatively 30\% were detected when consanguineous mating was allowed, with I$_{k}$ values of 
approximatively 20\%, to be compared with the average 30\% for the entire non- or moderately consanguineous population (not shown). 
This indicates that, as expected, the heterozygosity index I$_{k}$ of individuals decreases with the degree of consanguinity.
As a side note, when consanguinity was not allowed and with a starting population with variation level of 20\%, decreasing the initial 
crew members to 100 individuals had little effect on I$_{k}$ over the entire journey, as for 500 starting individuals, with 
I$_{k}$ remaining stable (around 30\%). With 30 people, however, the simulation terminated after 200 years because the consanguinity
threshold was reached rapidly. When consanguinity was allowed, 30 starting crew members produced descendants with lower I$_{k}$ 
(down to 22\%) and much higher consanguinity index (up to 30\%), highlighting that inbreeding and consanguinity occurred rapidly 
and that it accordingly, and as expected, decreased individuals’ heterozygosity index. With 100 or 500 starting people, the 
distribution of individuals’ I$_{k}$ was centered around 25-30\% and remained stable throughout the journey (seen Fig.~\ref{Fig:Individual_heterozygosity}),
indicating that 100 to 500 starting crew members is enough to stabilize I$_{k}$, as it was the case for polymorphism and 
H$_{i}$, i.e. to preserve allelic diversity and the proportion of heterozygous individuals.

\end{document}